\documentclass[]{interact}

\usepackage{epstopdf}
\usepackage[caption=false]{subfig}

\usepackage[numbers,sort&compress]{natbib}
\usepackage{amsmath,amssymb}
\bibpunct[, ]{[}{]}{,}{n}{,}{,}
\makeatletter
\def\NAT@def@citea{\def\@citea{\NAT@separator}}
\makeatother

\theoremstyle{plain}

\theoremstyle{definition}

\theoremstyle{remark}

\usepackage{color}

\begin{document}


\title{Effect of applied magnetic fields on the morphology of nematic nanobridges in slit pores}

\author{
\name{Pablo~Romero-Llorente\textsuperscript{a} and Jos\'e~Manuel~Romero-Enrique\textsuperscript{b,c}\thanks{Corresponding author: J.~M. Romero-Enrique. Email: enrome@us.es}}
\affil{\textsuperscript{a}Departamento de Sistemas F\'{\i}sicos, Qu\'{\i}micos y Naturales, Universidad Pablo de Olavide, Ctra. Utrera, km. 1 41013 Seville, Spain; \textsuperscript{b}Departamento de F\'{\i}sica At\'omica, Molecular y
Nuclear, \'Area de F\'{\i}sica Te\'orica, Universidad de Sevilla,
Avenida de Reina Mercedes s/n, 41012 Seville, Spain;  \textsuperscript{c}Instituto Carlos I de F\'{\i}sica Te\'orica y Computacional, Campus Universitario Fuentenueva,
Calle Dr. Severo Ochoa, 18071 Granada, Spain}
}

\maketitle

\begin{abstract}
In this paper we report a molecular dynamics study of the effect of the application of magnetic fields on the morphology of nematic nanobridges of $32000$ oblate Gay-Berne particles in slit pores favouring homeotropic anchoring. In absence of magnetic fields, previous studies show that there are different conformations of the nanobridge, depending on the slit pore width $D$ and the wettability of the walls. Under the application of uniform magnetic field cycles, in which the intensity of the field is increased stepwise until a maximum value and then decreased at the same rate, switching between different nanobridge conformations can be observed if the magnetic field is applied in a perpendicular direction to the global nematic director of the nanobridge. However, there are situations in which the initial bridge conformation is recovered after the magnetic field application, indicating that the switching can only be observed if there are different locally stable nanobridge conformations under the same thermodynamic conditions. Moreover, magnetic fields can destabilize the nanobridge, leading to its breakdown into isolated nanodroplets attached to a single wall. 
\end{abstract}


\section{Introduction}

Nematic liquid crystals are fluids in which long-range orientational order is present \cite{DeGennes1993}. However, interplay between elastic distortions, surface tension and anchoring can easily frustrate the orientational order of nematic liquid crystals in the presence of boundary surfaces, such as walls or interfaces, leading in general to the emergence of topological defects \cite{Kleman1983}, which are regions where the orientational and/or positional order abruptly vanishes and alter the long-range order around them, generating structures which stable against perturbations. In nematics, the most frequent topological defects are point defects or hedgehogs, and disclination lines, where the defect cores are either points or lines, respectively. Both types of defects are characterized by their winding number or topological charge, which accounts for the number of times the nematic director field rotates in a counterclockwise way if we follow a closed continuous path which encloses the topological defect core. For hedgehogs, it is usually either +1 or -1, where the minus sign means that the nematic director rotates in a clockwise way. On the other hand, for disclination lines it is either +1/2 or -1/2, which means that the nematic director rotates $\pi$ radians, either counterclockwise or clockwise, respectively. The manipulation of the orientational ordering by external electric or magnetic fields is known since the seminal work by Fr\'eedericksz and Repiewa \cite{Freedericksz}, which is the basis of the nowadays liquid crystal displays technology. In many cases, there are locally stable conformations of the bounded or confined nematics which differ by the orientational ordering of the liquid crystal, such as zenithal bistable nematic devices with different surface relief structures \cite{Uche_2005,Uche_2006,Evans_2010,Dammone2012}. The application of electric or magnetic fields can switch between these conformations which, in many cases, have different optical properties. Due to this fact, these systems have been proposed as key ingredients for practical applications such as privacy windows and other electro-optic devices, in which liquid crystal droplets dispersed in a polymer matrix change their optical properties under the application of external fields \cite{Drzaic1995,Sutherland19941074,Bunning200083,Bowley20019,Jazbinsek20013831,Rudhardt20032610,Fernandez-Nieves2006}. 

We will address the study of capillary bridges confined in slit capillaries. For this purpose, the Gay-Berne model for nematogens \cite{Gay19813316} is considered. This model has been used to model realistic molecular systems. This potential is parametrized by $\kappa$, which accounts for the anisotropy in the repulsive part of the potential, $\kappa'$ for the anisotropy in the attractive part of the potential, and $\mu$ and $\nu$, which determine the orientational dependence of the potential depth. A more precise definition of these parameters will be given in the next section. For example, the Gay-Berne model with $(\kappa,\kappa',\mu,\nu)=(4.4,20,1,1)$ has been used to parametrize $p$-terphenyl molecular interactions \cite{Luckhurst1993233,Bates19997087} (which do not show experimentally liquid crystalline behaviour), and $(0.345, 0.2, 1, 2)$ for the nematogen triphenylene core discotic model \cite{Emerson1994113,Cienega-Cacerez20143171}. The bulk phase diagram of this model has been extensively studied for both calamitic \cite{DeMiguel19901223,DeMiguel1991593,Chalam1991357,deMiguel1991174,deMiguel1991405,DeMiguel19923813,Rull1995113,DeMiguel19964234,Brown19986685} and discotic \cite{Emerson1994113,Bates19966696,Caprion2003417031,Bellier-Castella20044847,Chakrabarti2008} cases. Here we consider an oblate Gay-Berne $(0.5,1,2,1)$, which shows homeotropic anchoring on the vapour-nematic interface, i.e. the direction associated to the \emph{shortest} molecular axis orients preferentially parallel to the normal to the interface \cite{Rull2017,Rull2017_2}. In addition, the wall-particle interactions are chosen to promote homeotropic anchoring on the wall. This choice is motivated by the experimental work reported in \cite{Ellis2018} about 5CB capillary bridges on air in a slit pore formed by parallel glass surfaces with homeotropic anchoring conditions,which  reports different nematic configurations in the capillary bridge by tuning its diameter-to-height aspect ratio and shape, which can be either barrel-like or hourglass-like. So, for large aspect-ratio bridges a ring defect located on the equatorial plane is observed, while a point-defect structure is observed for small aspect-ratio bridges. On the other hand, the defects are either hyperbolic for hourglass-like bridges or radial for barrel-like bridges. However, the results obtained from Monte Carlo simulations for our oblate Gay-Berne model \cite{Romero2023} disagree for small aspect ratios with experiments. So, instead of the point-defect structure, simulations show biaxial bridge textures in which an annular disclination is arranged in a vertical plane. This can be qualitatively rationalized by the different length scales involving simulations and experiments \cite{Romero2023}. 

In this paper we will study how the application of magnetic fields can affect the nanobridge conformations. For this purpose, we will start from the cases reported in Ref. \cite{Romero2023} and, in each case, we will explore how the application of an uniform magnetic field with intensity modulated gradually in time can affect the morphology of the nanobridge. We will restrict ourselves to situations in which the magnetic field is applied to a direction perpendicular to the global nematic director. The reason for this is that, for large aspect-ratio bridges the global nematic director is perpendicular to the walls, i.e. the $z$ axis, but for small aspect ratios is on the transversal $xy$ plane. Thus, switching between configurations may be expected if the magnetic field is applied in the $x$ direction for large aspect-ratio bridges and in the $z$ direction for small aspect-ratio bridges.

The paper is organized as follows. The potential model and computer simulation methods will be presented in Section \ref{section2}. Section \ref{section3} will describe the principal findings of our simulations. We end up the paper with a discussion of our results and main conclusions.

\section{Methodology}
\label{section2}

\subsection{Potential model}

Particles are subject to a pairwise potential energy, the so-called Gay-Berne potential \cite{Gay19813316}, which is a modification of the Lennard-Jones potential that has been extensively considered in the study of nematogen fluids. It takes the form
\begin{equation}
U_{ij} ({\bf r}_{ij},{\bf u}_i,{\bf u}_j)
= 4\varepsilon (\hat{\bf r}_{ij}, {\bf u}_i, {\bf u}_j)
\left[\rho_{ij}^{-12}-\rho_{ij}^{-6}\right],
\end{equation}
where
\begin{equation}
\rho_{ij} = \frac{ r_{ij} - \sigma (\hat{\bf r}_{ij},
{\bf u}_i, {\bf u}_j) + \sigma_0 }{\sigma_0}.
\end{equation}
Here ${\bf u}_{i}$ is the unit vector along the symmetry axis
of particle $i$, $r_{ij}=|{\bf r}_i - {\bf r}_j|$ is the distance
along the intermolecular vector $\bf{r}_{ij}$ joining the centers
of mass of particles $i$ and $j$,
and $\hat{\bf r}_{ij}={\bf r}_{ij}/r_{ij}$.
The anisotropic contact distance,
$\sigma(\hat{\bf r}_{ij},\bf{u}_i,\bf{u}_j)$, and the depth of the
interaction energy, $\varepsilon (\hat{\bf r}_{ij}, {\bf u}_i,
{\bf u}_j)$,
depend on the orientational unit vectors, the length-to-breadth ratio
of the particle, $\kappa = \sigma_{ee} / \sigma_{ss}$, and the
energy depth anisotropy, $\kappa'= \epsilon_{ee} / \epsilon_{ss}$
which are both defined as the ratio of the size and energy interaction
parameters in the end-to-end ($ee$) and side-by-side ($ss$)
configurations, respectively. With these definitions, $\kappa > 1$ 
corresponds to prolate particles while $\kappa < 1$ corresponds to oblate 
particles. Their expressions are given by
\begin{eqnarray}
\frac{\sigma (\hat{\bf r}_{ij}, {\bf u}_i, {\bf u}_j)}{\sigma_0}=
\Bigg[1-\frac{\chi}{2}\Bigg(\frac{(\hat{\bf r}_{ij}\cdot {\bf u}_i
+\hat{\bf r}_{ij}\cdot {\bf u}_j)^2}{1+\chi({\bf u}_i \cdot {\bf u}_j)}
+\frac{(\hat{\bf r}_{ij}\cdot {\bf u}_i
-\hat{\bf r}_{ij}\cdot {\bf u}_j)^2}{1-\chi({\bf u}_i \cdot {\bf u}_j)}
\Bigg)\Bigg]^{-1/2}
\end{eqnarray}
and
\begin{eqnarray}
\frac{\varepsilon (\hat{\bf r}_{ij}, {\bf u}_i,{\bf u}_j)}{\epsilon_0}=
[\epsilon_1({\bf u}_i,{\bf u}_j)]^{\nu}\times [\epsilon_2(\hat{\bf r}_{ij},
{\bf u}_i,{\bf u}_j)]^{\mu},
\end{eqnarray}
where
\begin{eqnarray}
\epsilon_1({\bf u}_i,{\bf u}_j)&=&
[1-\chi^2({\bf u}_i \cdot {\bf u}_j)^2]^{-1/2},
\\
\epsilon_2(\hat{\bf r}_{ij},{\bf u}_i,{\bf u}_j)&=&
1-\frac{\chi'}{2}\Bigg[\frac{(\hat{\bf r}_{ij}\cdot {\bf u}_i
+\hat{\bf r}_{ij}\cdot {\bf u}_j)^2}{1+\chi'({\bf u}_i \cdot {\bf u}_j)}
+\frac{(\hat{\bf r}_{ij}\cdot {\bf u}_i
-\hat{\bf r}_{ij}\cdot {\bf u}_j)^2}{1-\chi'({\bf u}_i \cdot {\bf u}_j)}
\Bigg],
\end{eqnarray}
$\chi = (\kappa^2 - 1) / (\kappa^2 +1)$ and $\chi'=[
(\kappa')^{1/\mu} - 1] / [(\kappa')^{1/\mu} + 1]$. 
Here $\sigma_0$ is the side-by-side 
intermolecular collision diameter, and $\epsilon_0$ is $[2\kappa/(\kappa^2+1)]^\nu$ times the minimum 
intermolecular potential energy between two molecules in the side-by-side 
configuration. We will consider the original Gay and Berne choice for $\mu=2$ and $\nu=1$  \cite{Gay19813316}. 
The length-to-breadth geometrical ratio $\kappa$ and
$\kappa'$ plays an important role in the formation of ordered
phases, and also determine the nematic phase anchoring on the nematic-vapor phase 
\cite{Mills19983284,Rull2017}. Thus, the nematic
anchors homeotropically to the nematic-vapor interface if $\kappa\le \kappa'$, being
random-planar otherwise. 

In addition to the fluid-fluid interactions, particles are subject to two external potentials. The first one is the interaction with the slit pore of width $D$ along the $z$ axis
\begin{equation}
U_{w,i}(z_i)=U_{GB,w}(z_i)+U_{GB,w}(D-z_i),
\label{U_wall}
\end{equation}
where $z_i$ is the height of the particle $i$ above the wall. The single-wall-particle potential $U_{w,i}$ is a $9-3$ potential:
\begin{equation}
U_{GB,w}(z_i)=a\frac{2\pi}{3}\rho_w\sigma_w^3 \epsilon_w\left[\frac{2}{15}\left(\frac{\sigma_w}{z_i}\right)^9-\left(\frac{\sigma_w}{z_i}\right)^3\right].
\label{U_wall_2}
\end{equation}
Here $\rho_w \sigma^3=0.988$, $\sigma_w/\sigma_0=1.096$, $\epsilon_w/\epsilon_0=1.277$, which are the parameters for the wall-particle interaction in the Ar-CO$_2$ system \cite{Finn1989}. Finally, $a$ is an adimensional wall interaction strength which can be varied. Although his potential does not depend on the orientation of the GB particles (unlike in Ref. \cite{Rull2007}), this wall promotes homeotropic anchoring to the nematic phase \cite{Romero2023}. 

Finally, the interaction of each particle with an uniform magnetic field $\mathbf{B}=B\mathbf{u}_B$ is given by
\begin{equation}
U_i^B =\frac{\lambda}{2}\left(1-3(\mathbf{u}_i\cdot \mathbf{u}_B)^2\right),
\end{equation}
where $\lambda$ is the magnetic interaction strength parameter, which has the expression \cite{Satoh2006}
\begin{equation}
\lambda=\frac{\Delta \chi B^2}{3\mu_0},
\label{deflambda}
\end{equation}
with $\mathbf{u}_i$ being the symmetry axis orientation, $\Delta \chi$ the magnetic polarizability and $\mu_0$ the vacuum magnetic permeability.
 $\Delta \chi$ the anisotropy in the molecular magnetic polarizability, which is usually positive for calamitic nematogens and negative for discotic nematogens \cite{Dunmur}, and $\mu_0$ the vacuum magnetic permeability. Despite in this paper a model for discotic liquid crystals is considered, we assume that $\Delta \chi$ is positive.
This potential favours particle to orient along the magnetic field or opposite to it by inducing a torque $\mathbf{T}_i^B$ given by
\begin{equation}
\mathbf{T}_i^B=3\lambda (\mathbf{u}_i\cdot \mathbf{u}_B) (\mathbf{u}_i\times \mathbf{u}_B).
\end{equation}

\subsection{Simulation procedure}

We carry our molecular dynamics simulations of a system of $N=32000$ GB oblate particles with $\kappa=0.5$ and $\kappa'=1$ by using the LAMMPS open source package \cite{LAMMPS,LAMMPS2,LAMMPS3} in an in-house-modified version to include the interaction between particles and the magnetic field. We used LAMMPS implementation of the GB potential \cite{Berardi,Perram}, which is unshifted, and we choose a cutoff radius of $4\sigma_0$. Energies, lengths and times are given in units of $\epsilon_0$, $\sigma_0$ and $\sigma_0\sqrt{m/\epsilon_0}$, respectively, where $m$ is the mass of the particle. Accordingly, the reduced temperature $T^*$ and number density $\rho^*$ are defined as
\begin{equation}
T^*=\frac{k_B T}{\epsilon_0}\qquad,\qquad \rho^*=\rho \sigma_0^3\ .
\label{reducedunits}
\end{equation}
Hereafter we will drop the asterisk, so quantities are given in reduced units. Previous Monte Carlo simulations \cite{Rull2017,Rull2017_2} showed that nematic-vapour coexistence appears in a narrow range of temperatures $T=0.4-0.5$. We choose $T=0.5$, in which the density of the nematic phase at coexistence was estimated as $\rho=1.86(2)$, with a nematic order parameter $S=0.50$ \cite{Rull2017,Rull2017_2}. In addition, the nematic-vapour interface shows homeotropic anchoring \cite{Rull2017_2}, as expected for $\kappa<\kappa'$.

The simulation box is a square cuboid of dimensions $L\times L \times D$, with $L=45$ and different values of $D$. Periodic boundary conditions were imposed in the $x$ and $y$ directions, while two walls are placed at $z=0$ and $z=D$. In this paper, we consider slit pores with $D=20$, $30$ and $40$, similarly to the cases considered in Ref. \cite{Romero2023}, and adimensional wall interactions strengths $a=1$ and $1.25$. Simulations are performed in the $NVT$ ensemble with the Nose-Hover thermostat \cite{Nose,Hoover} for a fixed temperature $T=0.5$. The time step is set to $\Delta t=0.002$ and the thermostat relaxation time to $\tau=0.01$.

The magnetic field application cycle is composed by two stages. In the first stage, we perform simulations in which the parameter $\lambda$, Eq. (\ref{deflambda}), is stepwise increased by $\Delta \lambda=0.025$ each $2\times 10^5$ steps from a null value up to a maximum value $\lambda_{max}$. Afterwards, during the second stage the values of $\lambda$ are stepwise decreased again by $\Delta \lambda=0.025$ each $2\times 10^5$ steps until the zero-magnetic field case is reached. This gradual modulation in time of the magnetic field intensity is chosen to allow the nanobridge to equilibrate in most stages of the magnetic field cycle. For each value of $D$, the starting configuration is the bridge configuration obtained from the Monte Carlo simulations for $a=1$, with velocities and angular velocities chosen randomly following the corresponding equilibrium Maxwell-Boltzmann distribution. As we wish to induce a switch between different orientational textures by the magnetic field application, magnetic fields favour different orientations from the initial ones. Thus, for $D=20$, in which the initial condition is axisymmetric about the $z$ axis, magnetic fields are applied along the $x$ axis with a maximum value $\lambda_{max}=0.1$. For this value, the bridge-averaged nematic order parameter $S$ is slightly above  the bulk value, and most particles are aligned to the magnetic field throughout the bridge. On the other hand, for $D=30$ and $D=40$,  in which a disclination line on a vertical plane is observed in the initial configuration, magnetic fields are applied along the $z$ axis with a maximum value $\lambda_{max}=0.15$. This maximum value was selected in a similar way to the case of a magnetic field applied along the $x$ axis.

For each value of $\lambda$, different physical quantities, such as the mean potential energy and the global orientational ordering, are monitored during the simulation. The instantaneous global nematic ordering is characterized by the tensor order parameter in each step:
\begin{equation}
\mathbf{Q}= \frac{1}{N}\sum_{i=1}^N \frac{3{\bf u}_i\otimes{\bf
u}_i-\textrm{\bf I}}{2} \label{defq}
\end{equation}
where $\mathbf{u}_i$ is the unit vector parallel to the symmetry axis of the particle $i$, $\otimes$ represents the tensorial product and $\textrm{\bf I}$ is the identity matrix.
Its largest eigenvalue, $S$, and its associated eigenvector $\mathbf{N}$, characterize the
instantaneous global nematic order. These properties help us to characterize the reorientation process when the magnetic field is applied. 

In addition, the shape and nematic texture within the bridge are characterized
by calculating density and orientational profiles. The density profile, $\rho(x,y,z)$, is obtained as:
\begin{equation}
\rho(x,y,z)\equiv \left\langle \sum_{i=1}^N \delta(x_i-x)\delta(y_i-y)\delta
(z_i-z) \right\rangle, \label{defrhoz}
\end{equation}
where $(x_i,y_i,z_i)$ are the Cartesian coordinates of the particle $i$, $\delta$ denotes Dirac's delta and $\langle \ldots \rangle$ stands for the time average, typically over the last $10^5$ steps of each simulation. On the other hand, the orientational order profile is
given by the local tensor order parameter
\begin{equation}
\mathbf{Q}(x,y,z)\equiv \frac{1}{\rho(x,y,z)}\left\langle \sum_{i=1}^N 
\frac{3{\bf u}_i\otimes{\bf u}_i-\textrm{\bf I}}{2} 
\delta(x_i-x)\delta(y_i-y)\delta(z_i-z)\right\rangle, \label{defqz}
\end{equation}
The scalar nematic order parameter profile $S(x,y,z)$ and the local director field $\mathbf{n}(x,y,z)$ are obtained as the largest eigenvalue and the corresponding eigenvector of $\mathbf{Q}(x,y,z)$, respectively. It is also possible to get the biaxiality profile as the difference between the two smallest eigenvalues, but we see that this quantity does not provide further information, so we will skip it in our analysis. For the numerical evaluation of the density profile and local tensor order parameter, the box is divided in a grid of cubic voxels of unit side length (in reduced units), so the density and local tensor order profiles are obtained by averaging in each voxel. This voxel size is chosen as a compromise between coarse-graining and statistical precision, since smaller sizes are subject to larger statistical fluctuations, while larger sizes lead to a blurred picture of the density and order parameter profiles.

\section{Results}
\label{section3}
\begin{figure}
\centering
\includegraphics{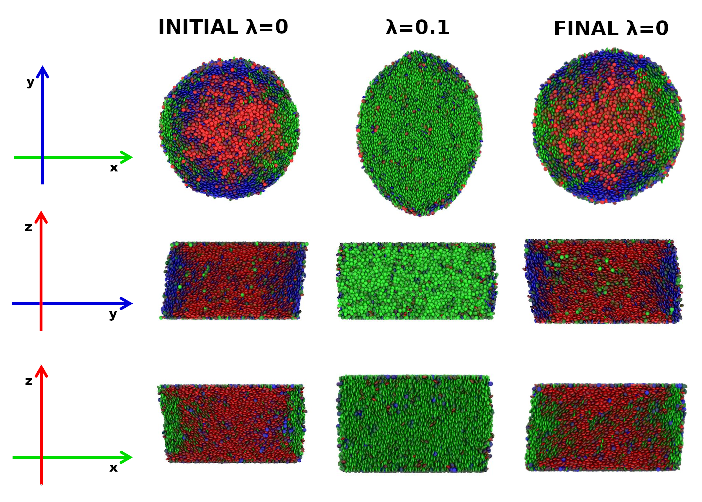} 
\caption{Snapshots of  $xy$, $yz$ and $xz$ cross sections of a bridge in a slit pore of two horizontal walls separated by a distance $D=20$ and $a=1$ at the end of the first $\lambda=0$ simulation, at $\lambda=0.1$ and at the end of the last $\lambda=0$ simulation of the magnetic field application cycle. The colour code associated to the GB particle orientations is the following: red if the main symmetry axes of the particles are aligned with the $z$ axis, blue if they are aligned with the $y$ axis and green if they are aligned with the $x$ axis.} \label{fig1}
\end{figure}
\subsection{$a=1$} 
We first considered slit pores in which the adimensional wall interaction strength $a=1$. Note that corresponds to the largest value considered in Ref. \cite{Romero2023} associated to hourglass-like bridges. Surprisingly, the first simulations at zero magnetic field for each $D$ change the shapes of the bridges, characterized by bigger contact angles but still hourglass-like, and in some cases lead to different orientational profiles. The explanation for these discrepancies may be related to the fact that, in the Monte Carlo simulations, the GB potential was shifted so takes zero value at the cutoff radius, while in the GB potential implemented in LAMMPS is unshifted. The shift does not affect the intermolecular forces, but it will alter the torques with respect to the unshifted version due to its orientation dependence. However, as we will see, we still observe a transition from a nearly axisymmetrical configuration where most the particles are oriented along the $z$ axis for $D=20$ to a configuration in which most particles are oriented along a transversal axis in the $xy$ plane for $D=30$ and $D=40$. So, we will explore if transitions between different bridge states can be obtained from the magnetic field application. 

\subsubsection{$D=20$}

\begin{figure}
\centering
\subfloat{%
\resizebox*{5cm}{!}{\includegraphics{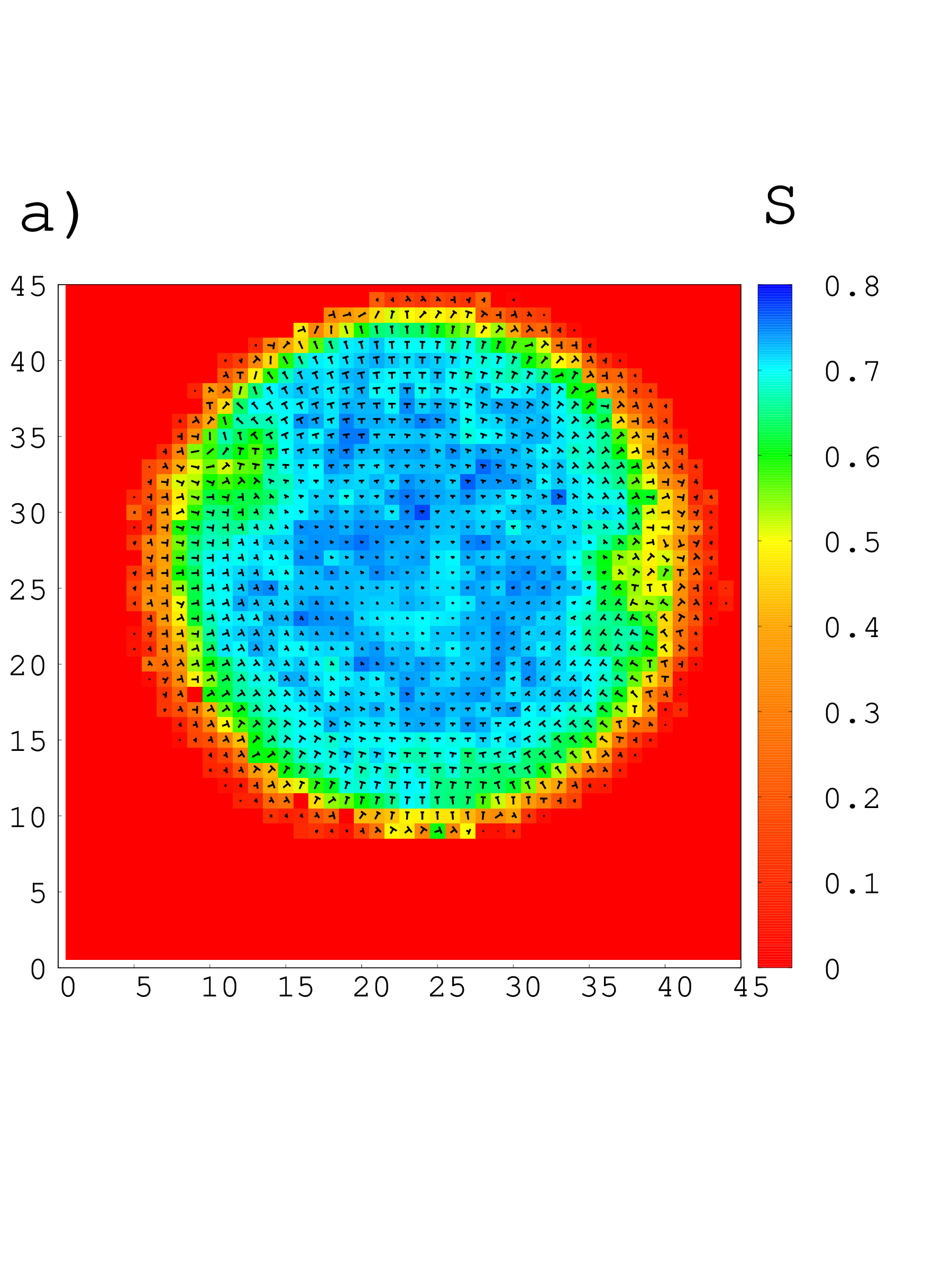}}}
\subfloat{%
\resizebox*{5cm}{!}{\includegraphics{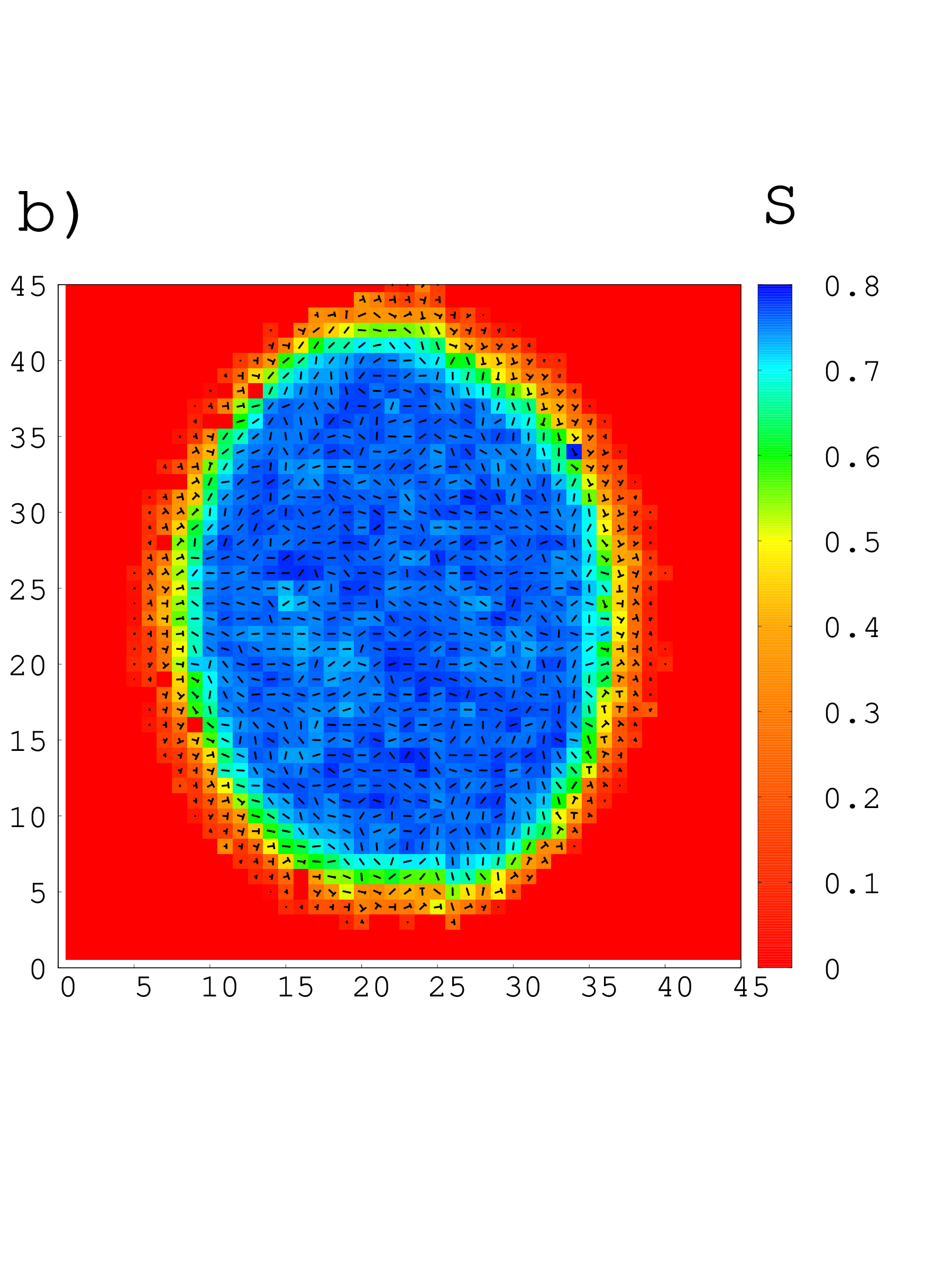}}}
\subfloat{%
\resizebox*{5cm}{!}{\includegraphics{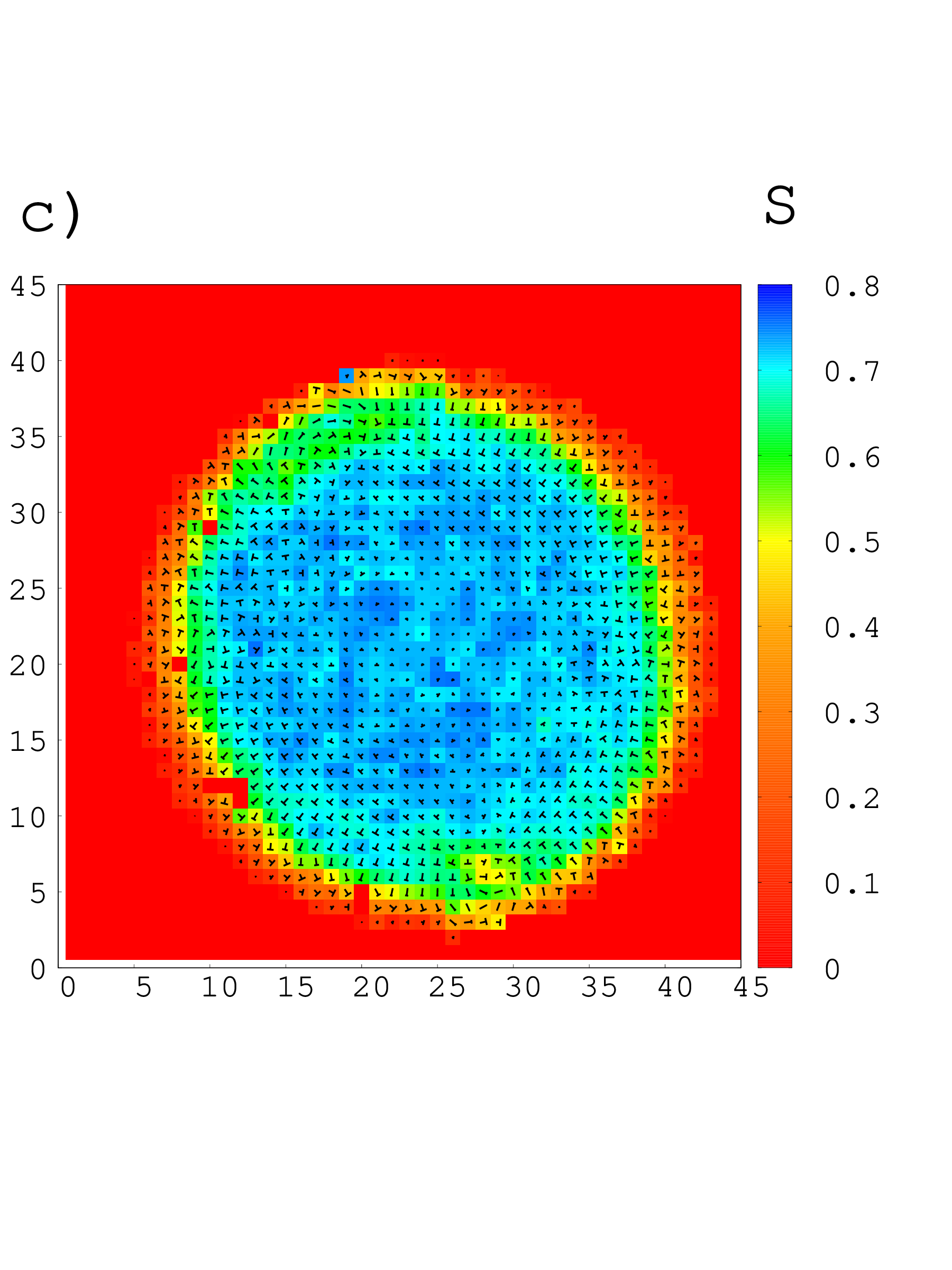}}}\vspace{-2.5cm}
\\
\subfloat{%
\resizebox*{5cm}{!}{\includegraphics{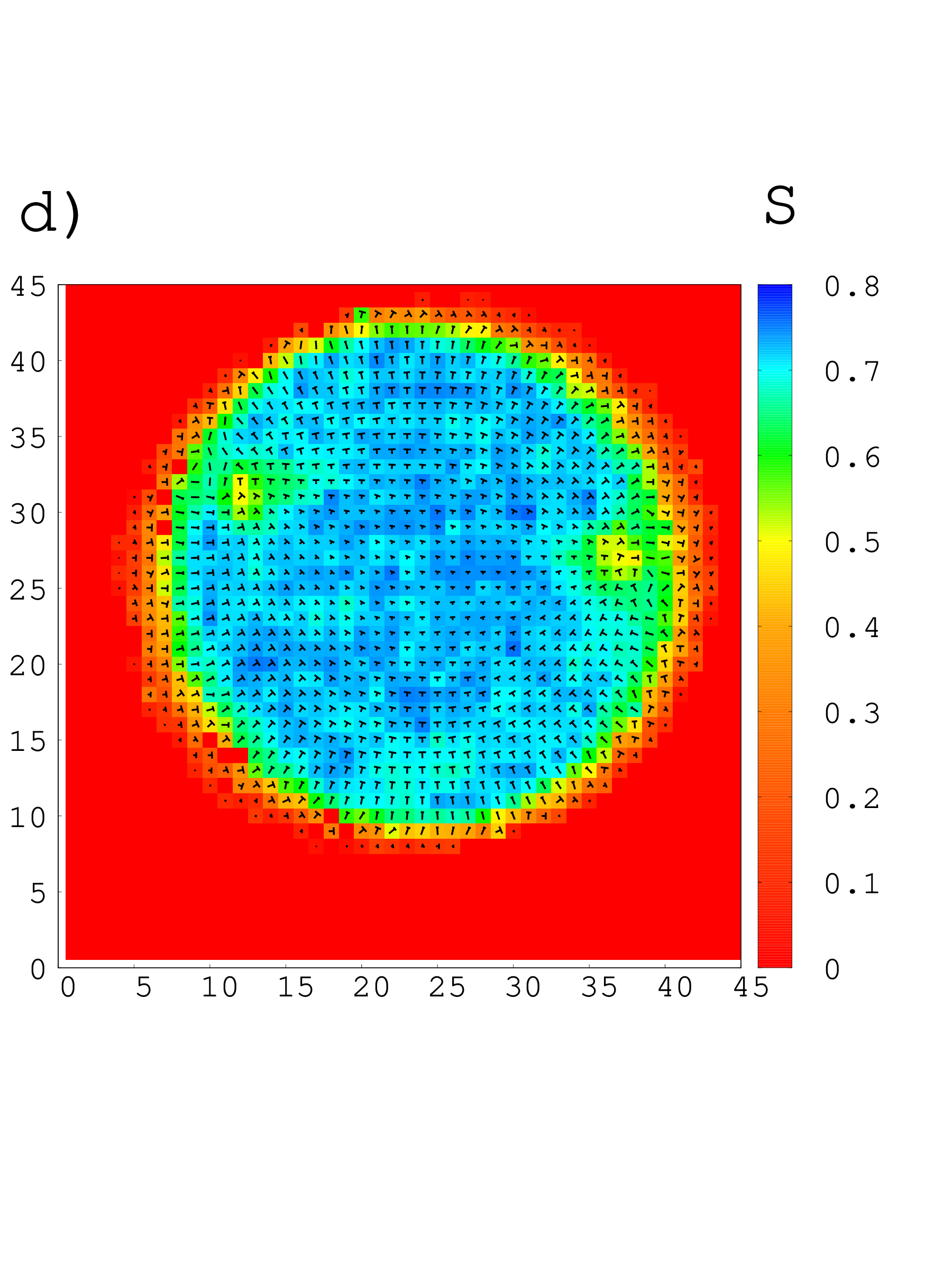}}}
\subfloat{%
\resizebox*{5cm}{!}{\includegraphics{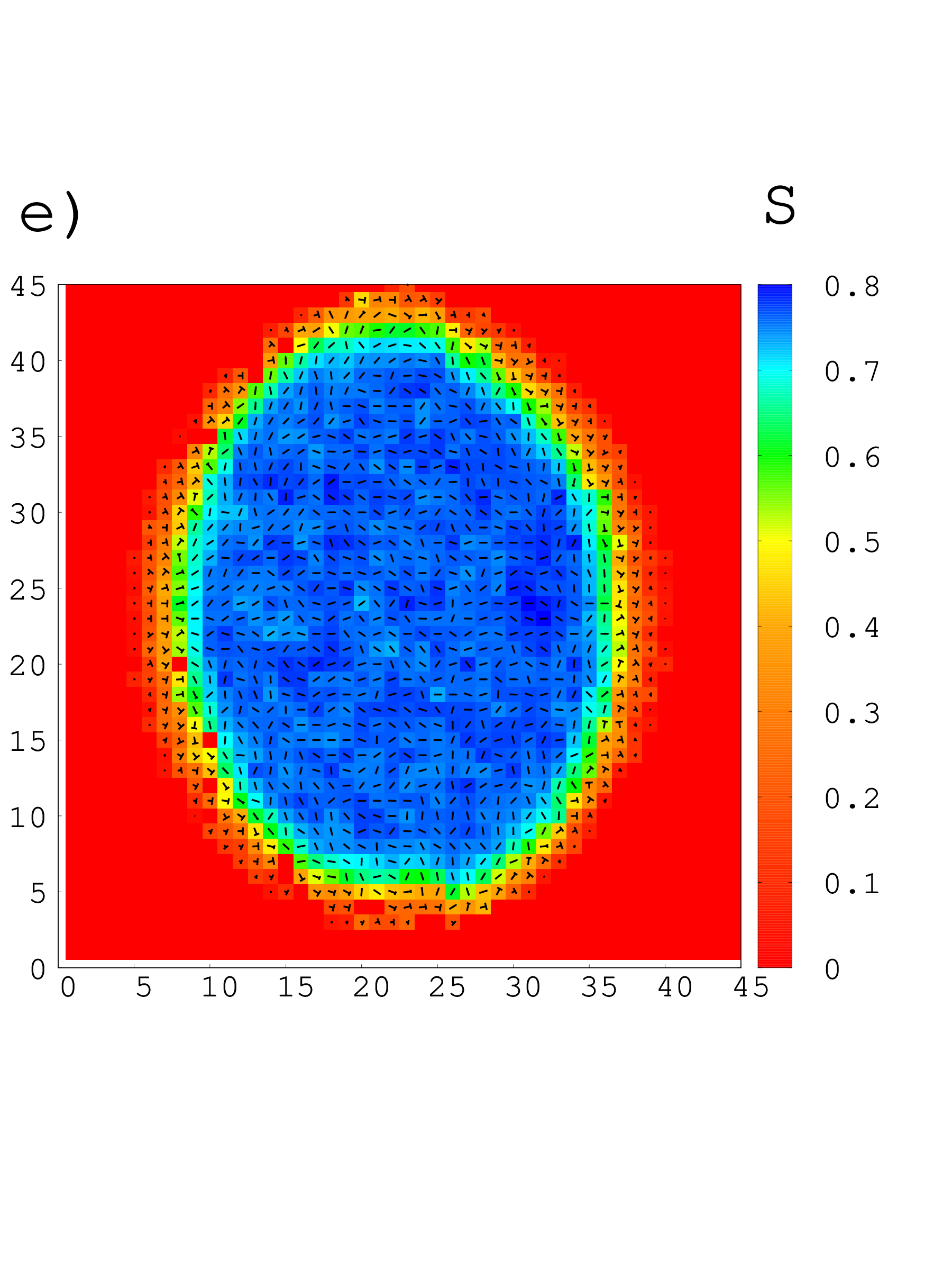}}}
\subfloat{%
\resizebox*{5cm}{!}{\includegraphics{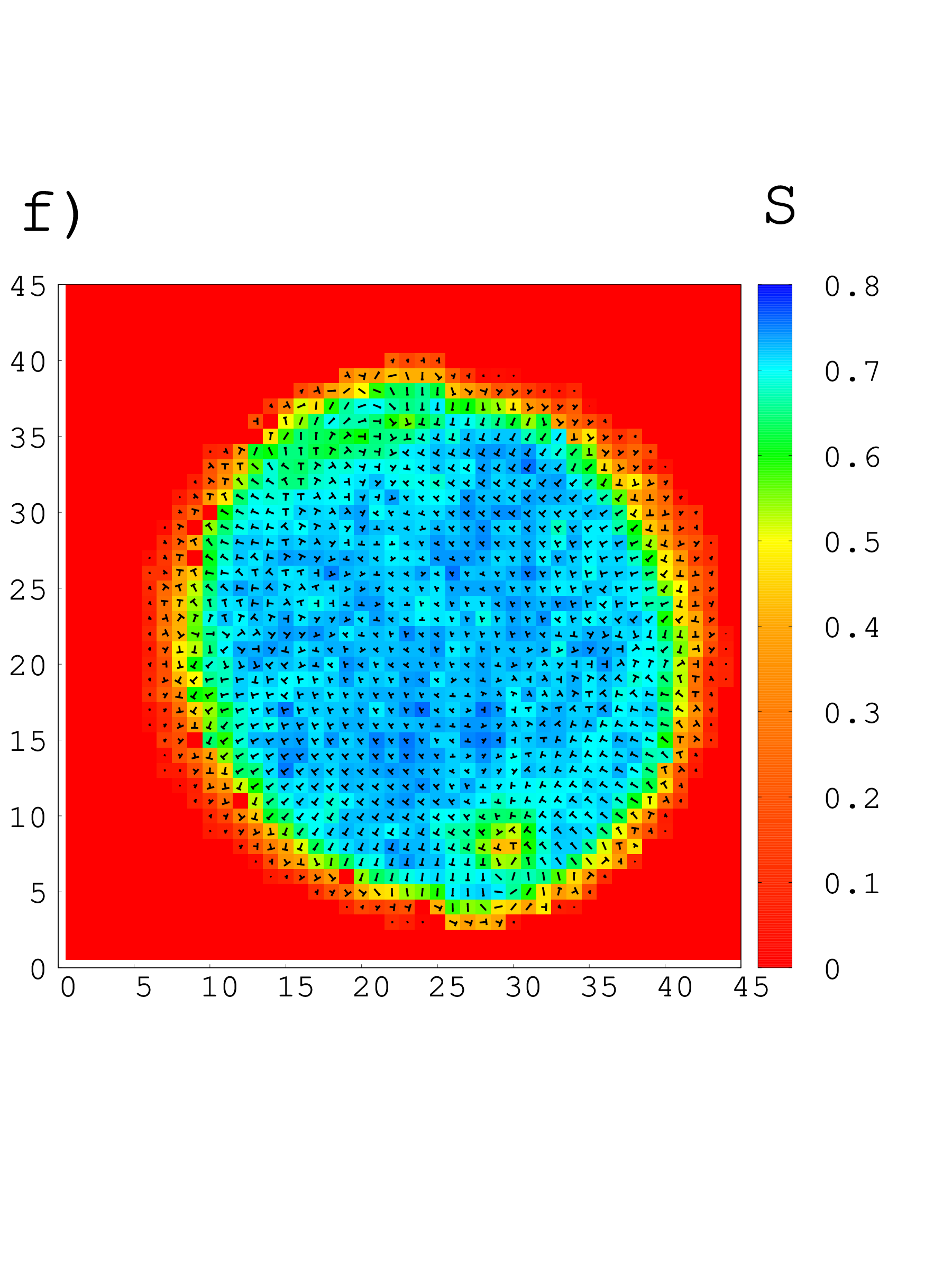}}}\vspace{-2.5cm}
\\
\subfloat{%
\resizebox*{5cm}{!}{\includegraphics{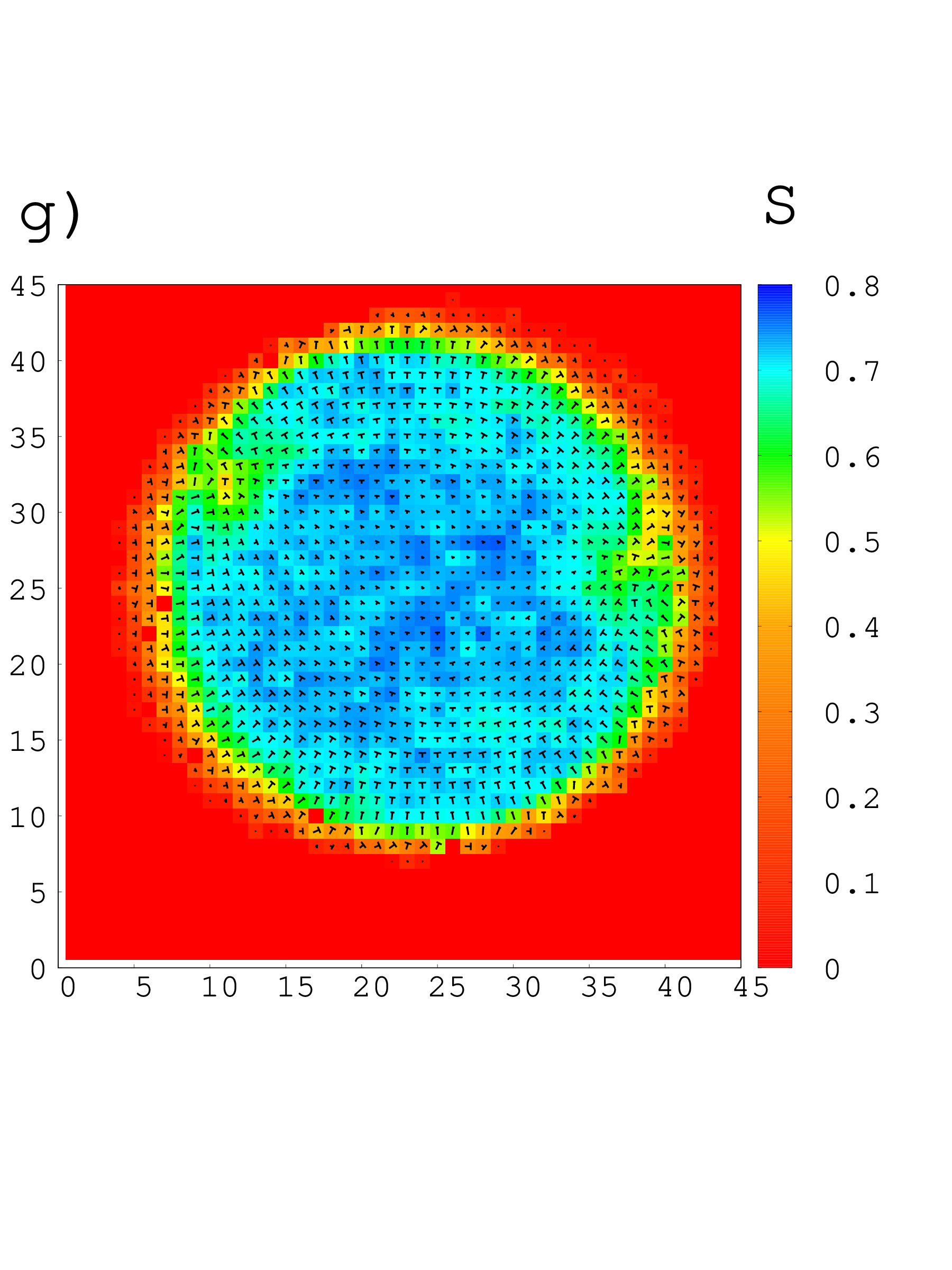}}}
\subfloat{%
\resizebox*{5cm}{!}{\includegraphics{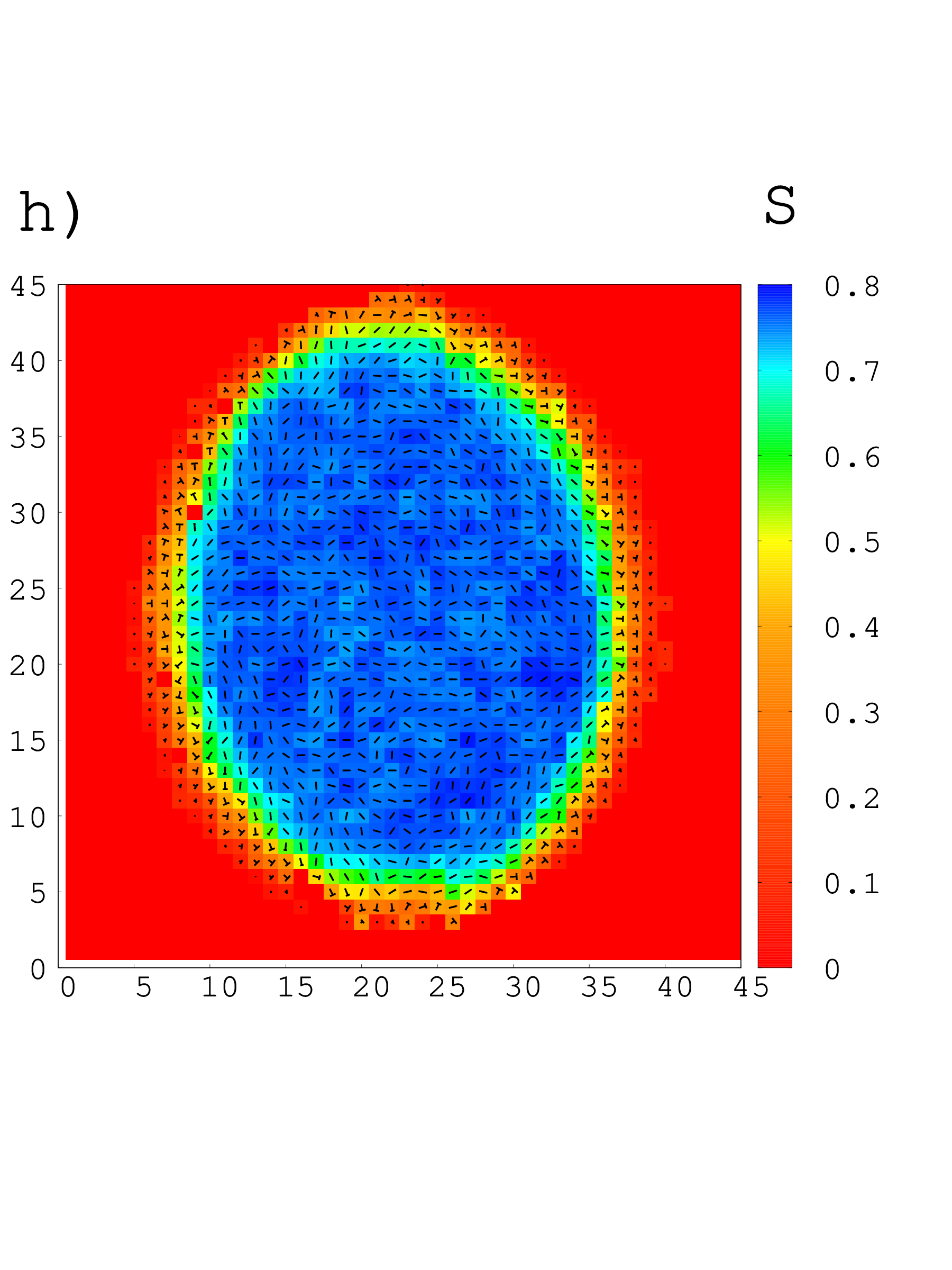}}}
\subfloat{%
\resizebox*{5cm}{!}{\includegraphics{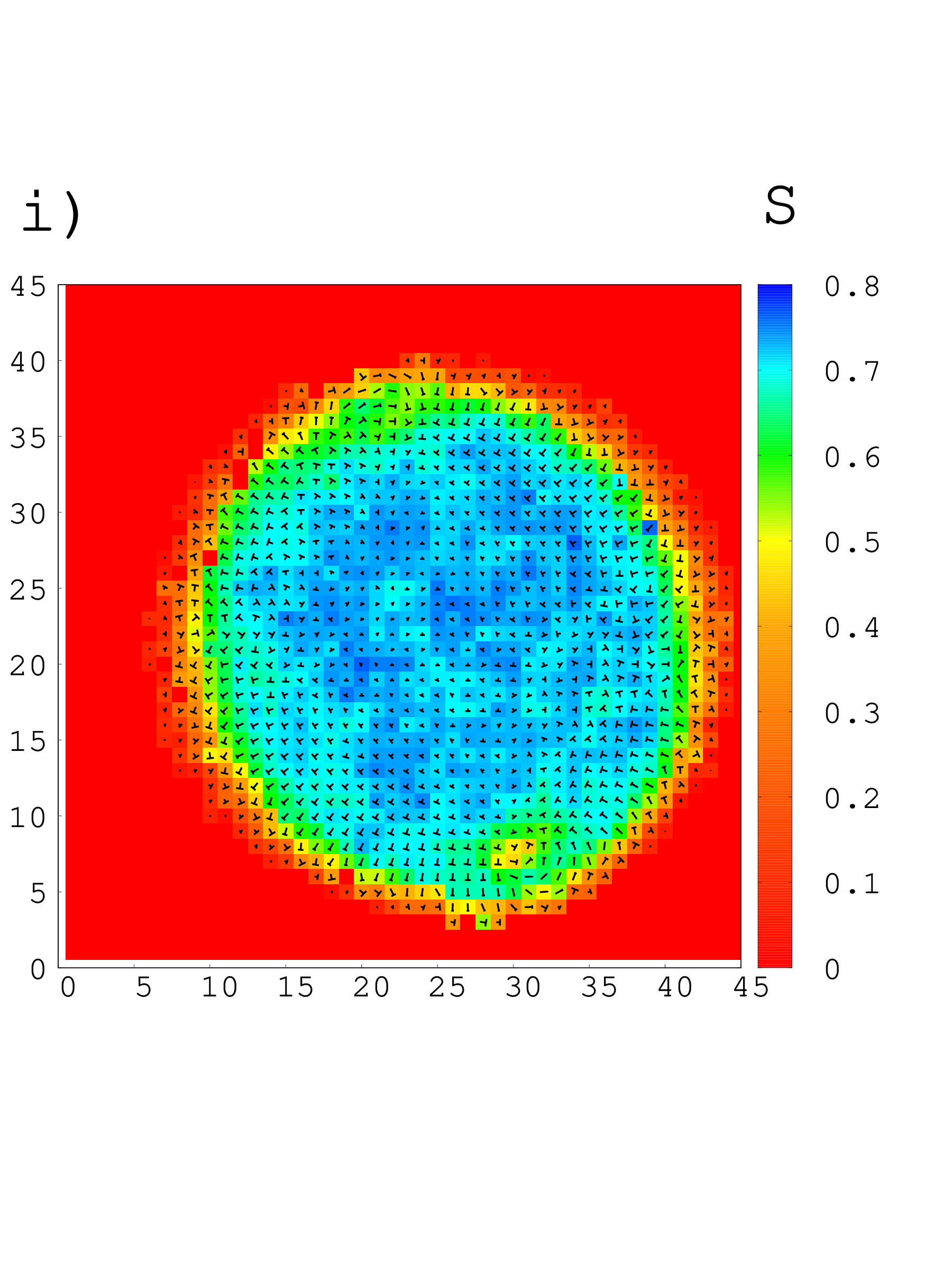}}}
\caption{Plots of the nematic order parameter $S$ profile (colour map) and nematic director field (nail representation, where the nail is parallel to the projection of tilted nematic director in the figure plane, and a perpendicular segment is plotted on the extreme of the nail with the lowest component of the director in the direction normal to the figure plane) of a cross section of the bridge for $a=1$ and $D=20$ at heights $z=5$ (upper row), $z=10$ (middle row) and $z=15$ (lower row). Left column corresponds to the initial $\lambda=0$ simulation, central column to the $\lambda=0.1$ simulation and right column to the final $\lambda=0$ simulation.} \label{fig2}
\end{figure}

We start with the case $D=20$. We consider in this case an $x-$direction magnetic field application cycle. Fig. \ref{fig1} shows different cross sections of a bridge at the final configurations of the first $\lambda=0$ simulation of the cycle, at the maximum value $\lambda=0.1$ and the last $\lambda=0$ simulation of the cycle. In the first zero-field configuration, we observe that the bridge is slightly sheared (not perfectly axisymmetric). Most of the particles orient parallel to the $z$ axis, except in the interfacial region, in which they preferentially orient homeotropically to the interface. In this region, the nematic order parameter $S$ smoothly decays from the bulk value inside the bridge core to zero outside the bridge, as well as the density (not shown by simplicity). This is a common feature of all the systems considered in this work. These findings qualitatively agree with the results reported for Monte Carlo simulation in Ref. \cite{Romero2023}. When the magnetic field is turned on, particles orient quickly along the $x$ axis (in fact, at the end of the $\lambda=0.025$ simulation most of the particles are already oriented along the $x$ axis). But, additionaly, the bridge is stretched along the $y$ axis (the larger the value of $\lambda$ is, the larger the stretching). This effect may be explained by the fact that this arrangement maximizes the surface perpendicular to the magnetic field. Such conformation optimizes the alignment of particles along the $x$ axis, reducing the magnetic contribution to the energy.  Finally, the configurations corresponding to the final $\lambda=0$ simulation is quite similar to the one from the initial $\lambda=0$ simulation, indicating that, under these conditions, the magnetic field cycle does not induce a transition to a different nanobridge conformation. 

Fig. \ref{fig2} plots the order parameter $S$ profiles and nematic director fields corresponding to selected cross sections (equatorial plane $z=10$, as well as intermediate heights $z=5$ and $z=15$) of the bridge for the initial and final $\lambda=0$ simulations, as well as the simulation for $\lambda_{max}=0.1$. For the first $\lambda=0$ simulation, the nematic director field confirms our previous observation that most particles orient along the $z$ axis in the central part of the bridge but mostly homeotropically on the interfacial region. However, there is no indication of a disclination ring located on the equatorial plane as it was reported in Ref. \cite{Romero2023}, but instead a smooth crossover compatible with an escaped radial texture \cite{Cladis,Meyer}. In the equatorial plane there is indication of two $+1/2$ vertical disclinations lines located in nearly opposite positions close to the interface, where the nematic order parameter $S$ is significantly reduced with respect to the bulk value and the nematic director rotates $+\pi$ radians in a counterclock closed loop around the disclination core. The $z=5$ and $z=15$ cross sections indicate that both disclinations emerge from the interface.
For $\lambda=0.1$, the nematic order parameter $S$ is uniform up to the interfacial region, with no indications of disclination lines. Note that the $S$ values are larger than these correspoding to the $\lambda=0$ cases. On the other hand, it is confirmed that the bridge shape is elongated along the $y$ direction while the nematic director is oriented mainly on the $x$ axis throughout the bridge. Finally, the last $\lambda=0$ simulation of the cycle shows a nematic order parameter profile and nematic director field very similar to the corresponding one in the initial $\lambda=0$ simulation. This strongly suggests that, after the cycle, the bridge returns to the initial state. Thus, the magnetic field cycle has not induced any transition in the bridge configuration for this case.

\subsubsection{$D=30$ and $D=40$}

\begin{figure}
\centering
\includegraphics{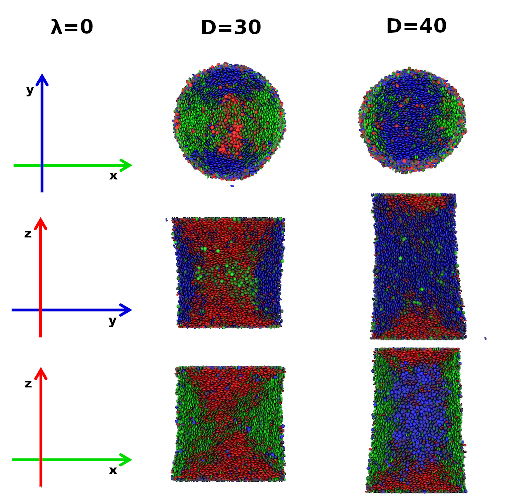} 
\caption{Final snapshots of the $xy$, $yz$ and $xz$ cross sections of a bridge in a slit pore of two horizontal walls with $a=1$ and separated by a distance $D=30$ (left column) and $D=40$ (right column) after a $\lambda=0$ simulation with $2\times 10^5$ steps. The colour code associated to particle orientations is the same as in Fig. \ref{fig1}.} \label{fig3}
\end{figure}

\begin{figure}
\centering
\includegraphics{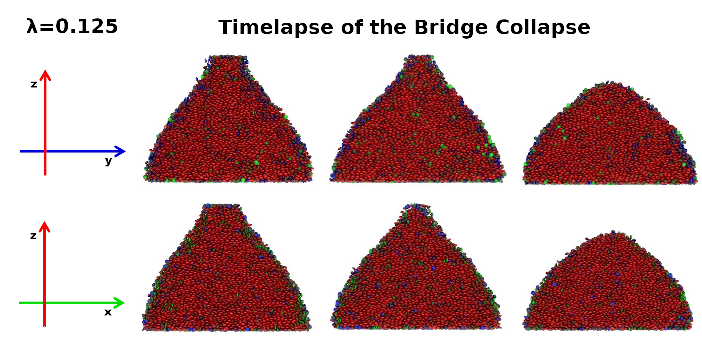} 
\caption{Time evolution of the bridge collapse for $D=30$, $a=1$ and $\lambda=0.125$. The $xz$ and $yz$ central sections of the bridge snapshots are shown for the time step $t=10^5$ (left column), $t=1.2\times 10^5$ (middle column) and $t=1.4\times 10^5$ (right column).
} 
\label{fig4}
\end{figure}

We turn now to the $D=30$ and $D=40$ cases. Similarly to the $D=20$ case, the initial $\lambda=0$ simulation from an initial condition obtained from Monte Carlo simulations deform the bridges, but they have similar features to those reported in Ref. \cite{Romero2023}, as it can be seen in Fig.~\ref{fig3}. Under the application of a magnetic field along the $z$ direction, particles reorient vertically, but in addition it amplifies any top-bottom asymmetry that the bridge may present, so now the contact area becomes smaller in one wall and larger in the other. If the magnetic field is strong enough, the bridge detaches from the wall where its contact surface is smaller and a single droplet is formed. Fig. \ref{fig4} shows this detachment for the $D=30$ case and $\lambda=0.125$ when the magnetic field intensity starts to decrease. A similar instability is observed for $D=40$. This process is irreversible and cannot be reverted by switching off the magnetic field. 
\subsection{$a=1.25$}
Now we increased the wall-particles interaction parameter to $a=1.25$. In this case, bridges have lower contact angles with the walls. As in the previous value of $a$, we analyze the effect of the magnetic field application on bridges with $D=20$, $D=30$ and $D=40$.
\subsubsection{$D=20$}
\begin{figure}
\centering
\includegraphics[width=\textwidth]{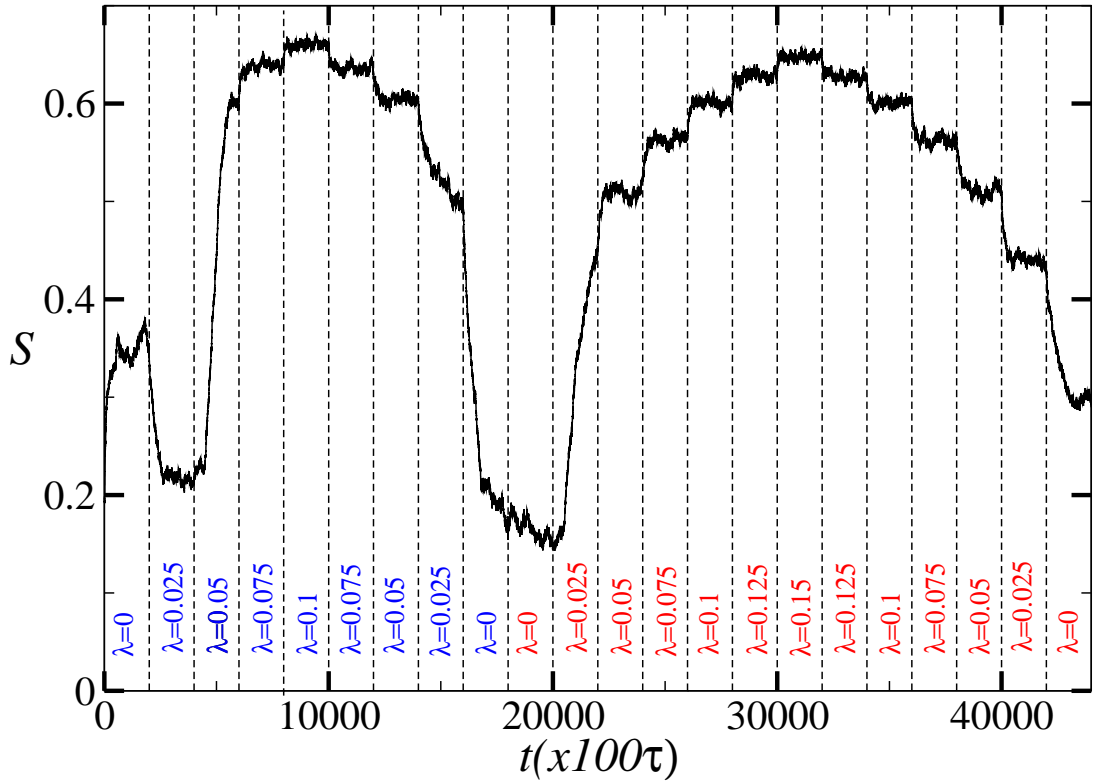} 
\caption{Time evolution of the instantaneous global nematic order parameter $S$ along the magnetic field application cycle for $D=20$ and $a=1.25$. For simplicity, the values are represented each $10^2$ steps. The values of $\lambda$ in blue correspond to the application of the magnetic field along the $x$ axis. On the other hand, red values of $\lambda$ represent magnetic fields along the $z$ axis.} 
\label{fig5}
\end{figure}
\begin{figure}
\centering
\includegraphics{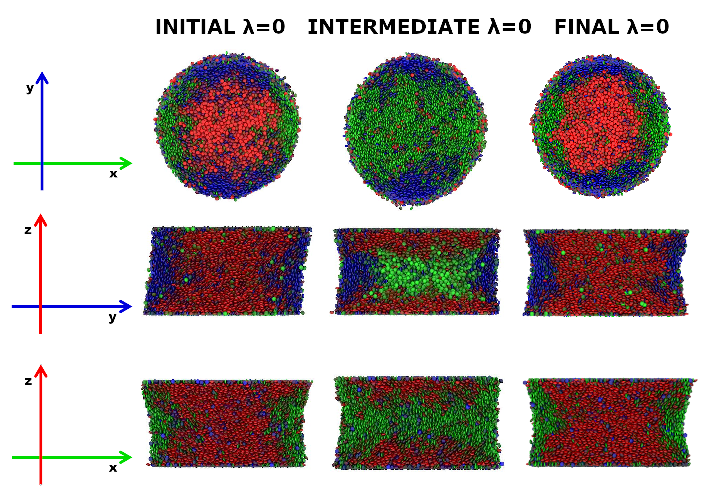} 
\caption{Snapshots of the $xy$, $yz$ and $xz$ cross sections of a bridge in a slit pore of two horizontal walls separated by a distance $D=20$ and $a=1.25$. The left column corresponds to the final configuration of the initial $\lambda=0$ simulation, middle column to the final configuration of the intermediate $\lambda=0$ simulation after the application of magnetic fields along the $x$ axis, and the right column to the final configuration of the last $\lambda=0$ simulation after the application of magnetic fields along the $z$ axis. The colour code associated to particle orientations is the same as in Fig. \ref{fig1}.} \label{fig6}
\end{figure}
\begin{figure}
\centering
\subfloat{%
\resizebox*{5cm}{!}{\includegraphics{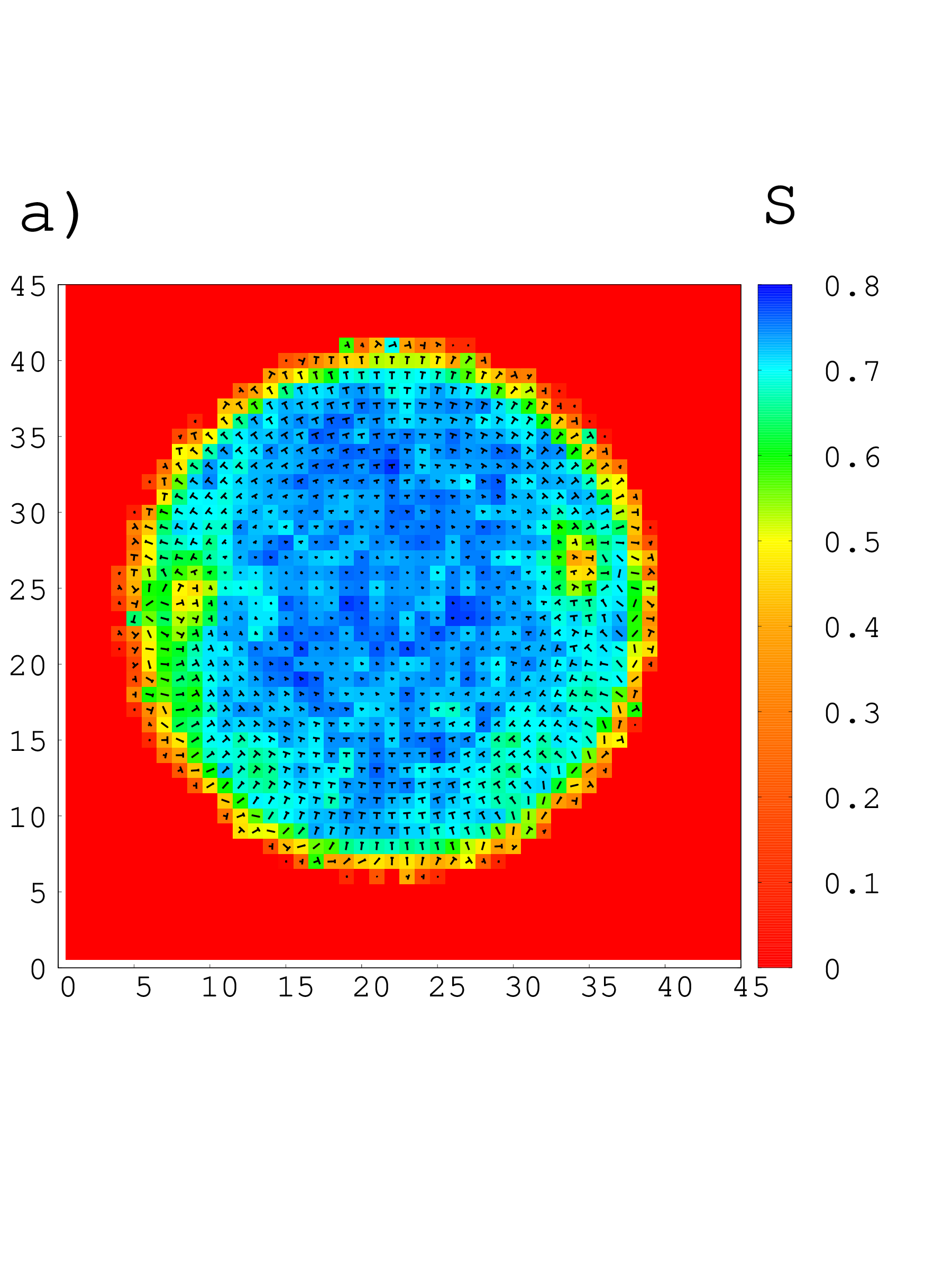}}}
\subfloat{%
\resizebox*{5cm}{!}{\includegraphics{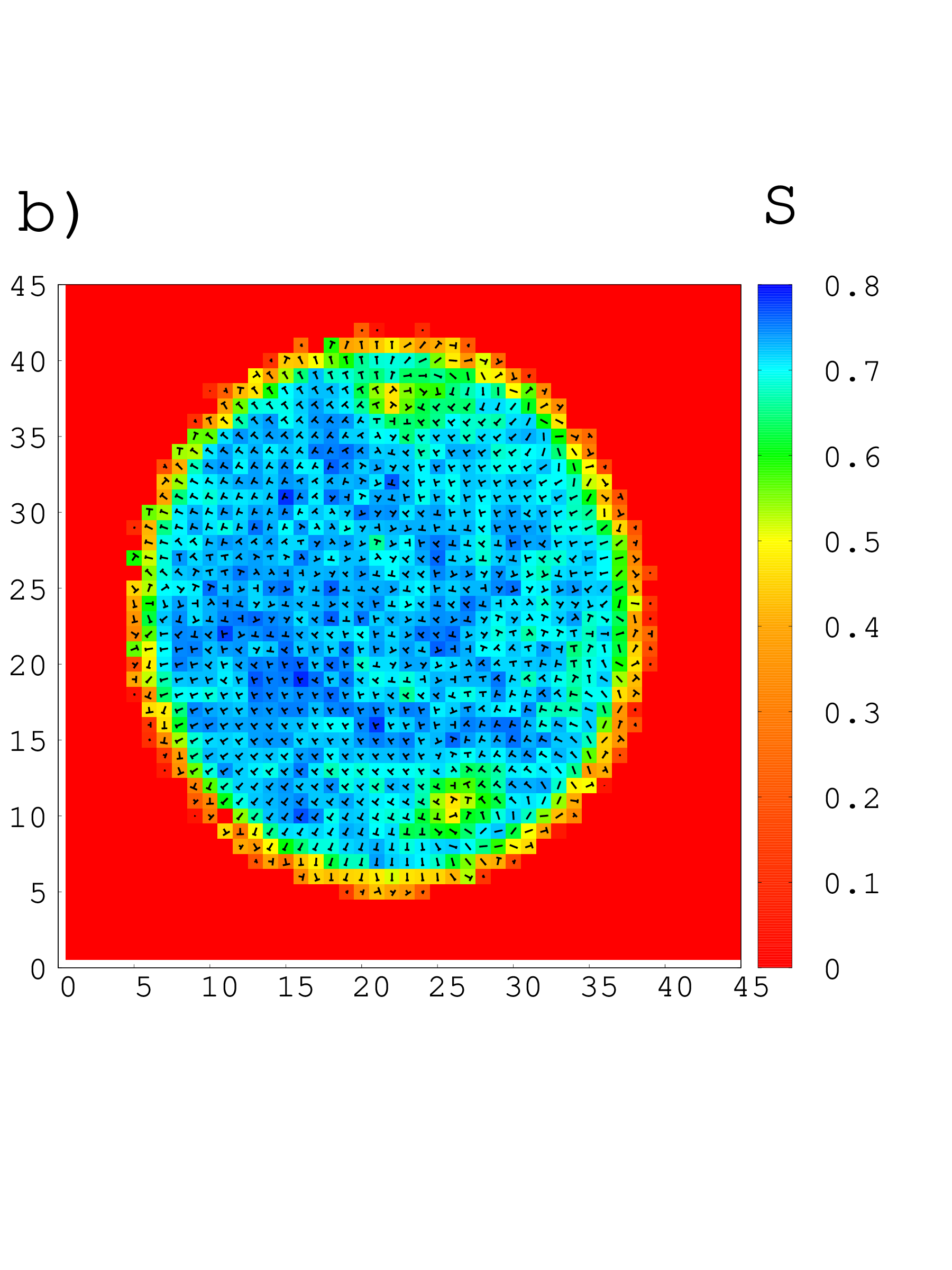}}}
\subfloat{%
\resizebox*{5cm}{!}{\includegraphics{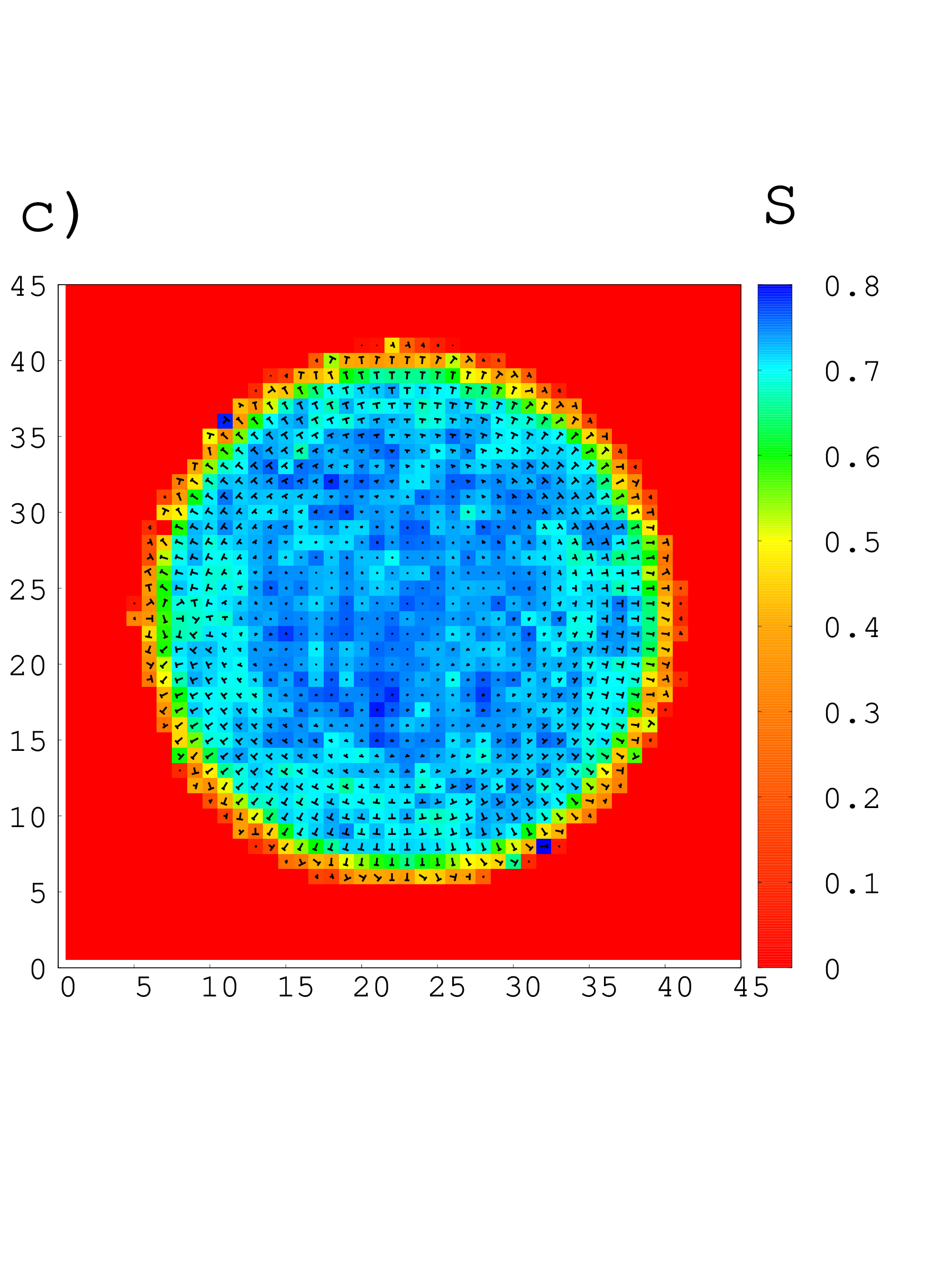}}}\vspace{-2.5cm}
\\
\subfloat{%
\resizebox*{5cm}{!}{\includegraphics{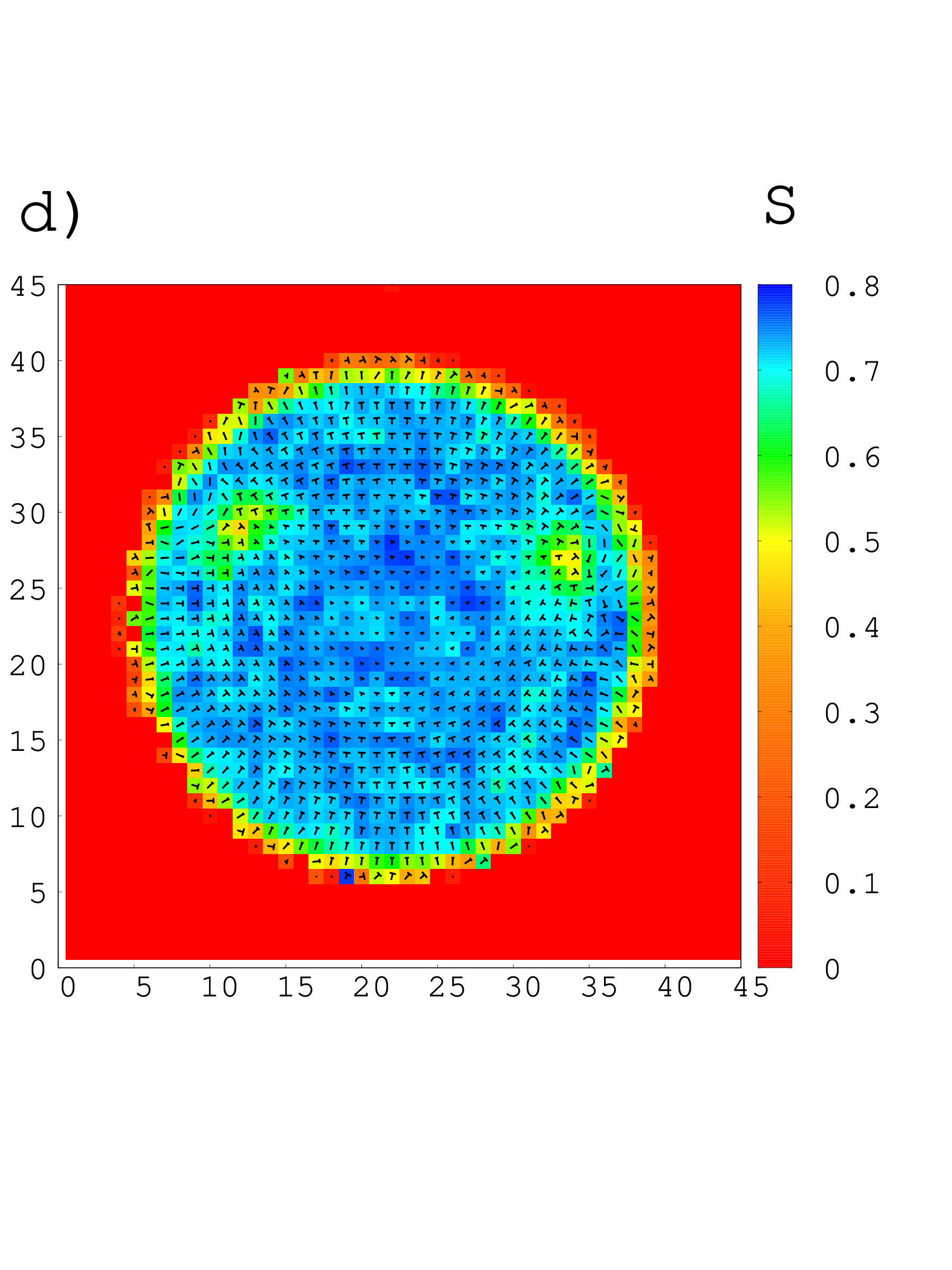}}}
\subfloat{%
\resizebox*{5cm}{!}{\includegraphics{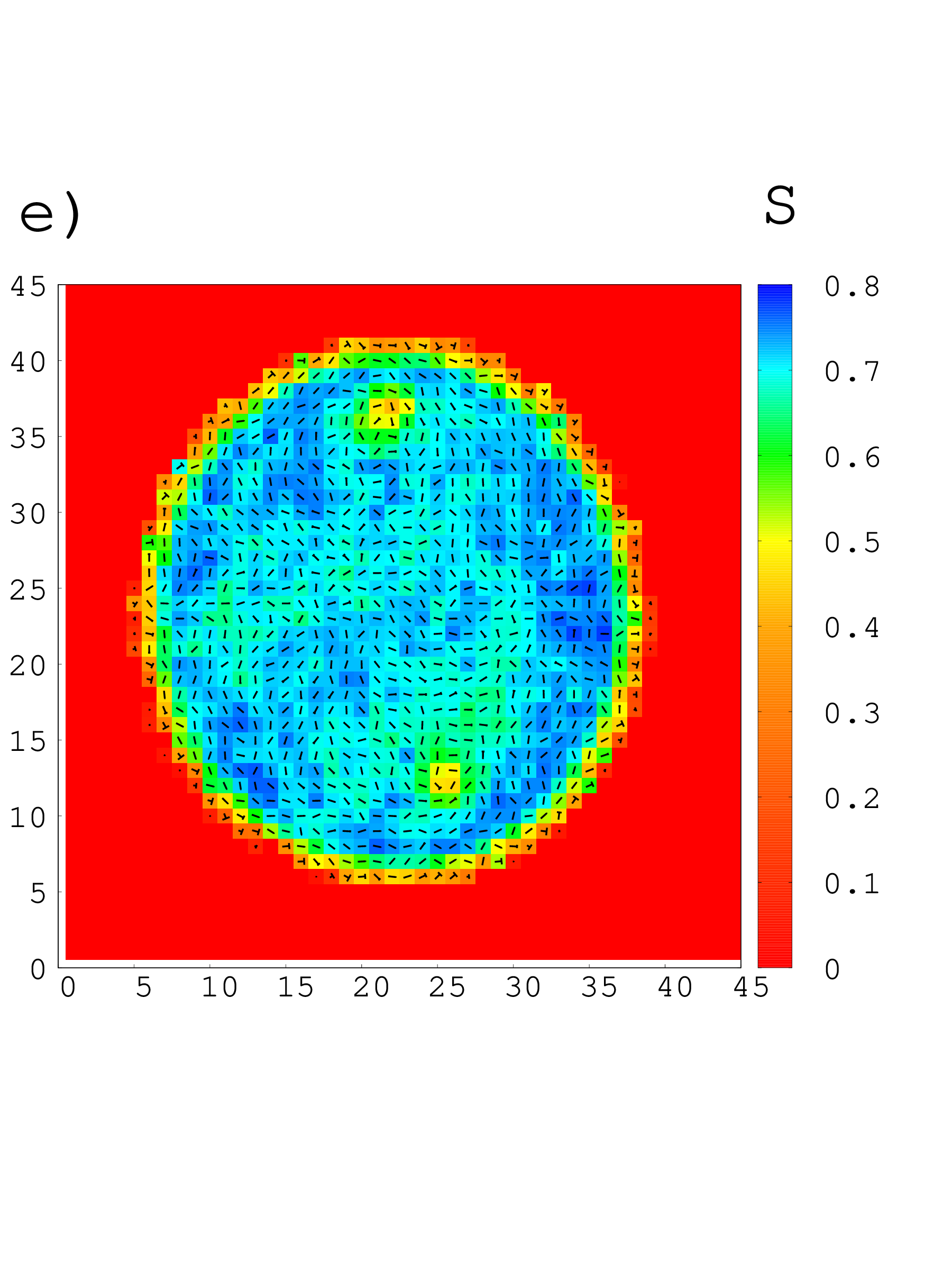}}}
\subfloat{%
\resizebox*{5cm}{!}{\includegraphics{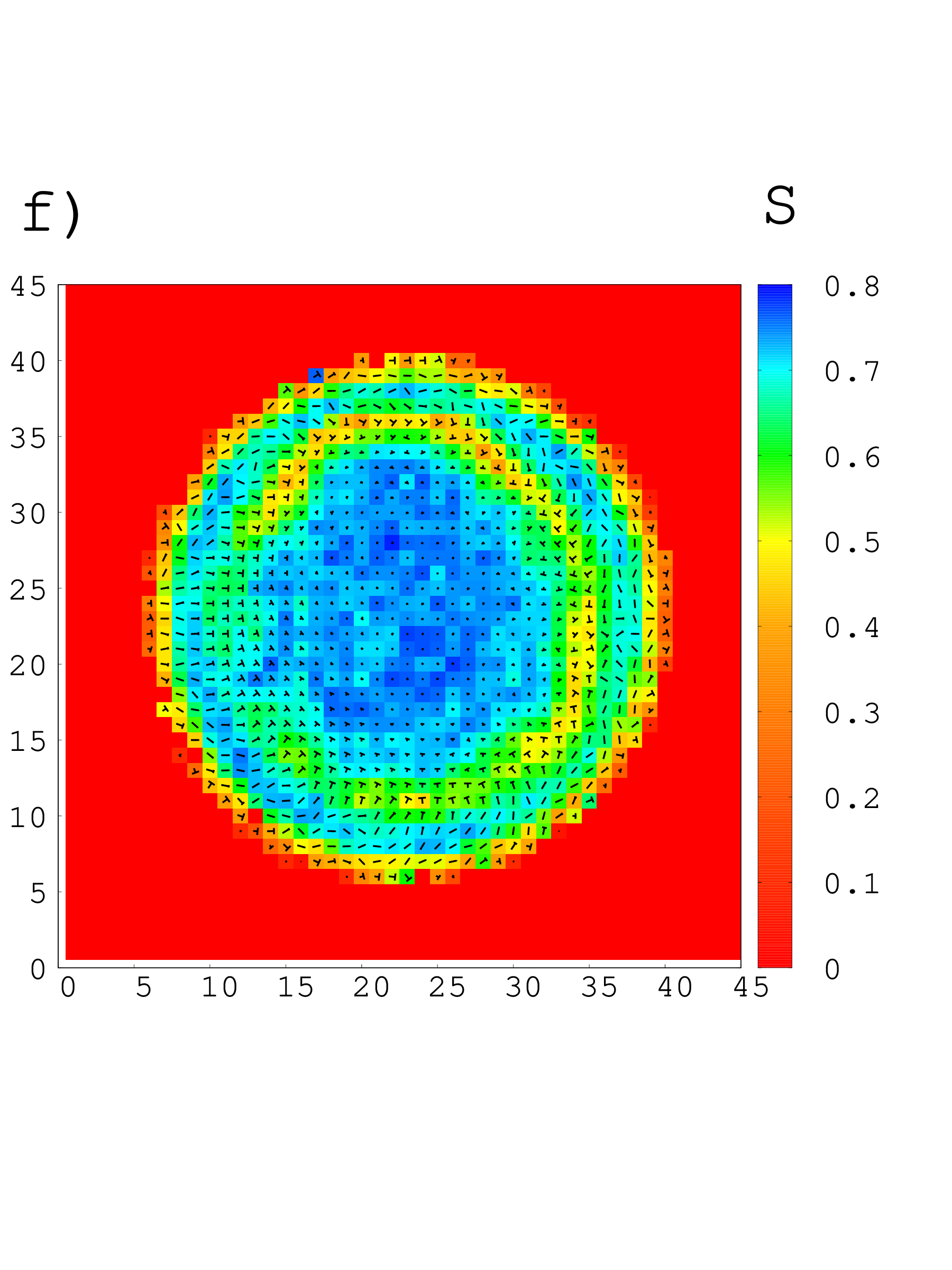}}}\vspace{-2.5cm}
\\
\subfloat{%
\resizebox*{5cm}{!}{\includegraphics{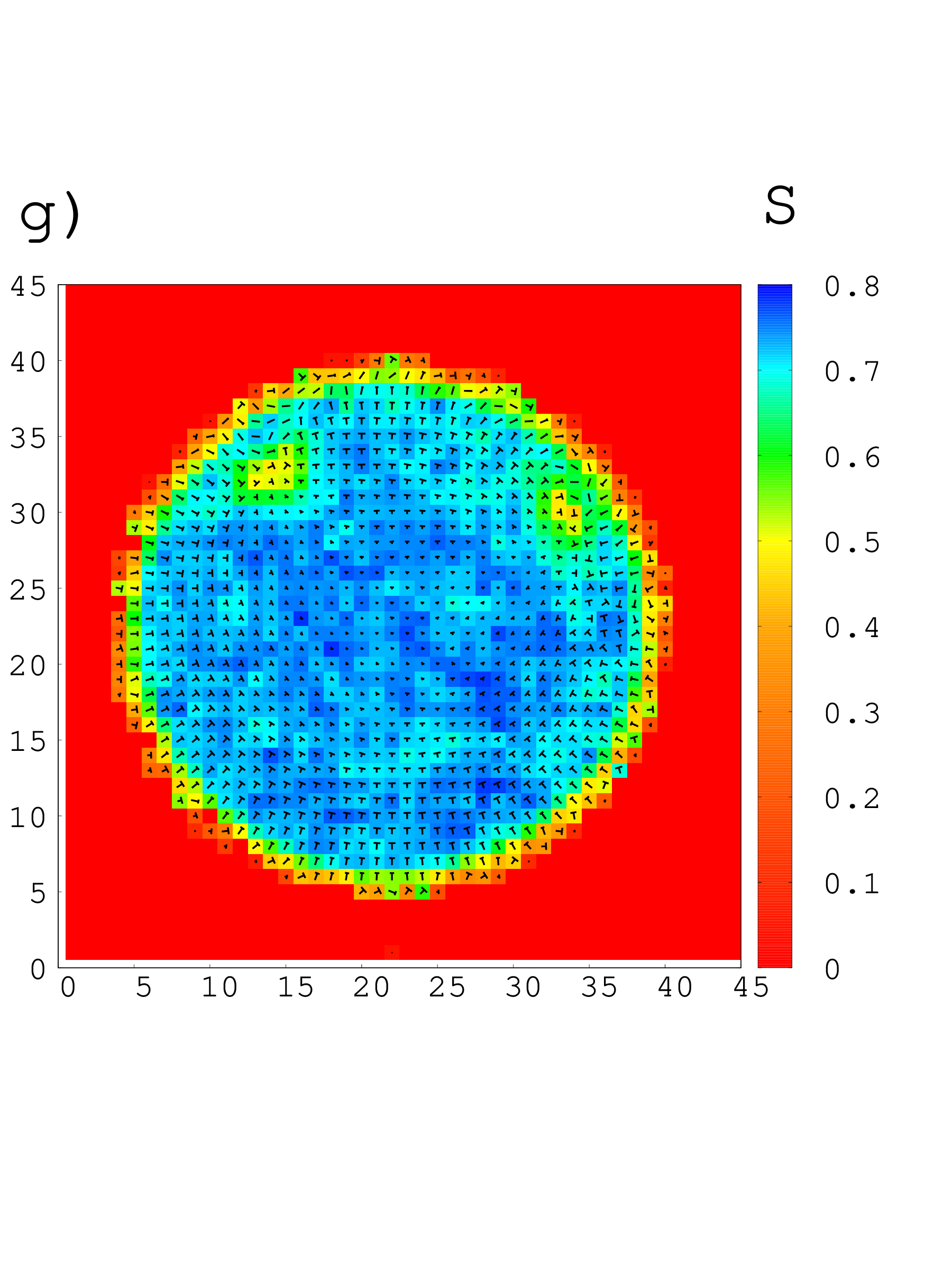}}}
\subfloat{%
\resizebox*{5cm}{!}{\includegraphics{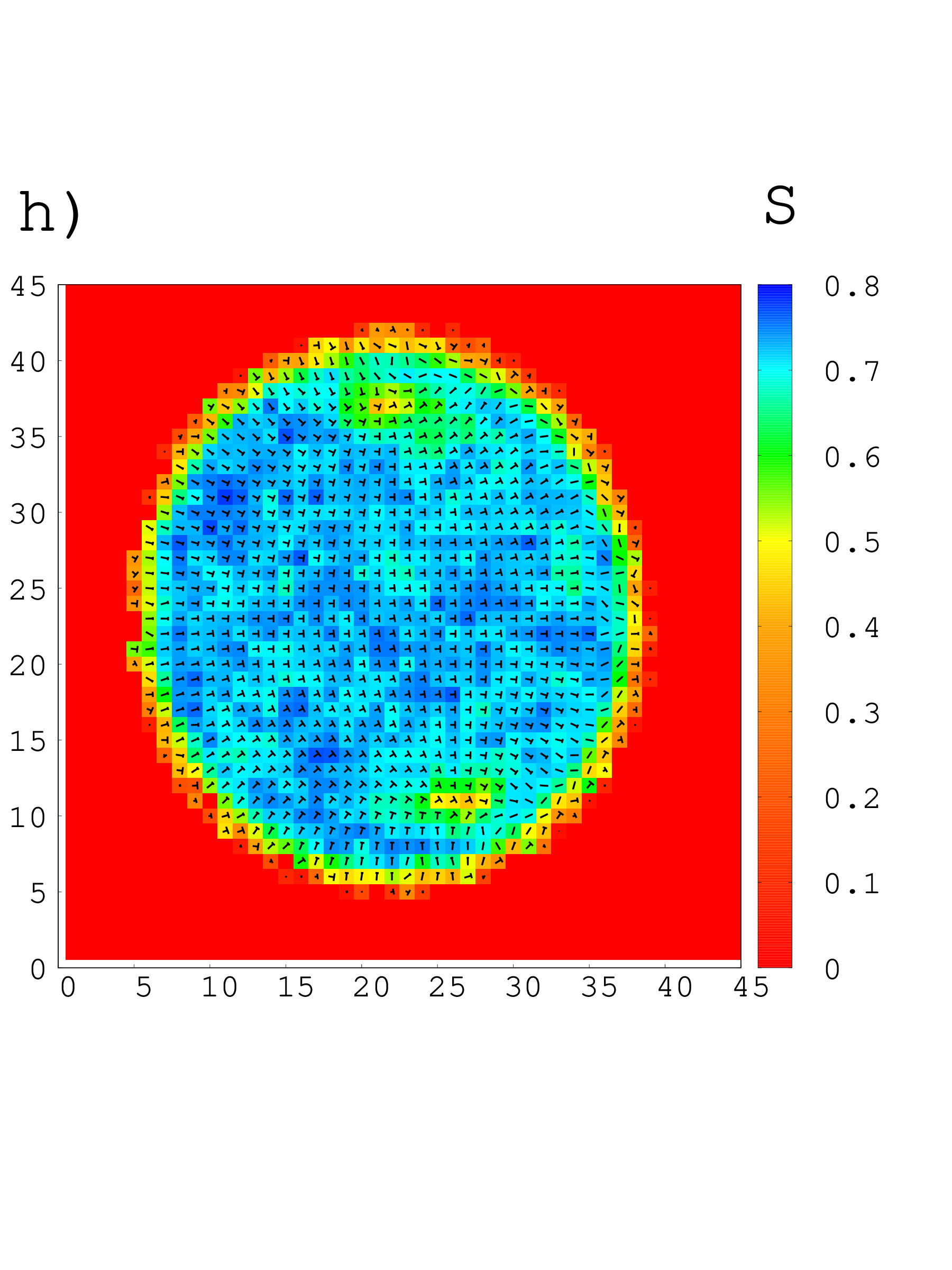}}}
\subfloat{%
\resizebox*{5cm}{!}{\includegraphics{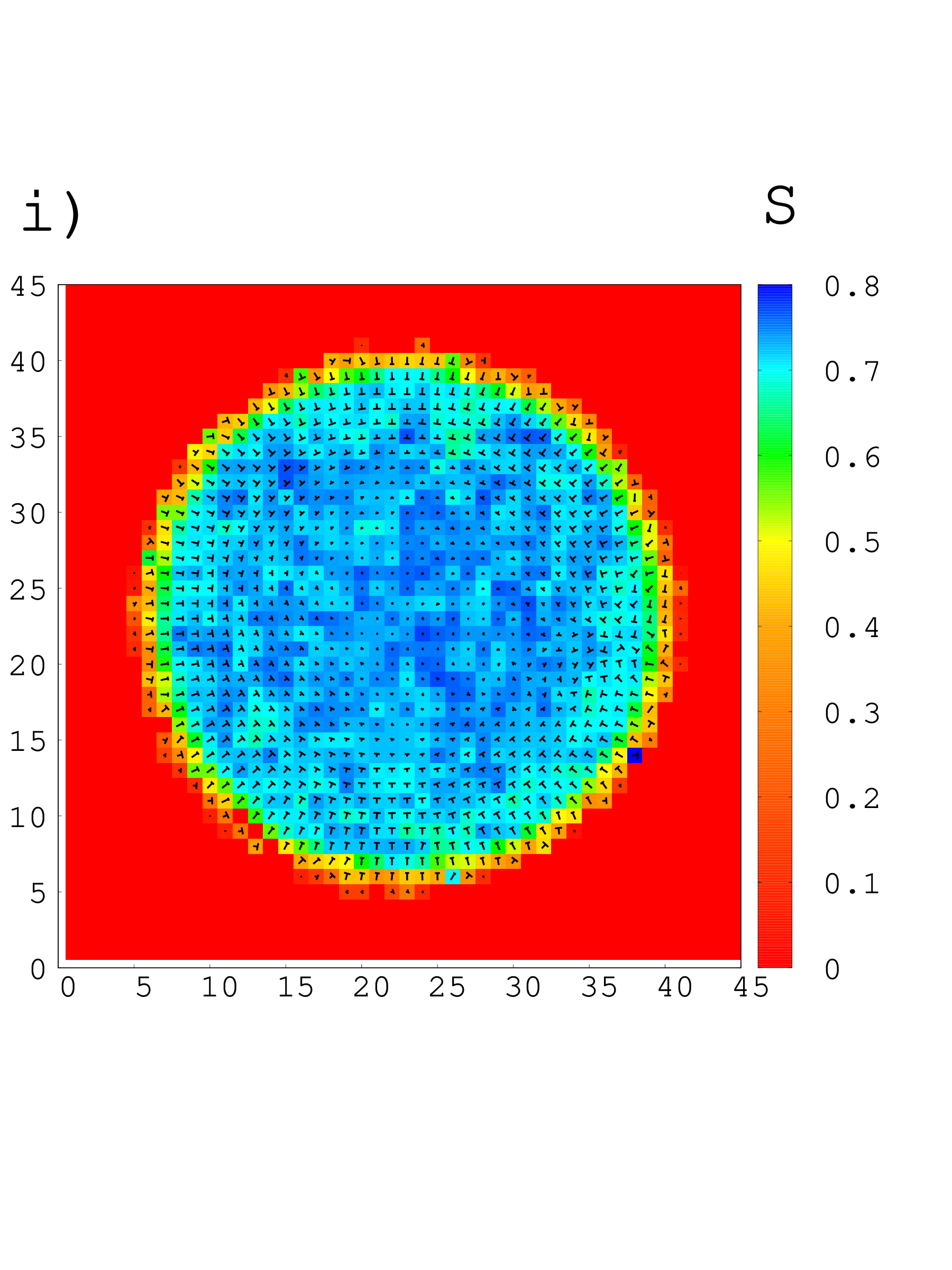}}}
\caption{Plots of the nematic order parameter $S$ profile (colour map) and nematic director field (nail representation) of a cross section of the bridge for $a=1.25$ and $D=20$ at heights $z=5$ (upper row), $z=10$ (middle row) and $z=15$ (lower row). Left column corresponds to the initial $\lambda=0$ simulation, central column to the intermediate $\lambda=0$ simulation and right column to the final $\lambda=0$ simulation.} \label{fig7}
\end{figure}

Fig. \ref{fig5} shows the evolution of the instantaneous nematic order parameter $S$ in a magnetic field application cycle. In a first stage, we followed a similar procedure to the $a=1$ case, in which the magnetic field is applied along the $x$ axis. The initial $\lambda=0$ simulation leads to a similar configuration to the obtained for the $a=1$ case (see left columns of Figs.~\ref{fig6} and \ref{fig7}). As in the previous case, the bridge is slightly sheared  and most of the particles are parallel to the $z$ axis but homeotropically-oriented in the interfacial region. Again, no indication of a disclination ring is observed, but instead a smooth crossover compatible with an escaped radial texture. In addition, two vertical disclinations emerge from the interface. As the magnetic field is turned on, the system evolves in a similar way as the $a=1$ case: particles orient preferentially along the $x$ axis and the bridge stretches along the $y$ direction (not shown by simplicity). It is worth noticing that $S$ changes abruptly at the beginning of the simulations for $\lambda=0.025$ and $\lambda=0.05$, indicating major particle reorientations in the bridge as the region of particles oriented along the $x$ axis grows from the interface neighbourhood to the center of the bridge. However, as the value of $\lambda$ is reduced to $0$, we observe that the final stage differs from the initial one, as it can be seen at the middle columns of Figs. \ref{fig6} and \ref{fig7}. After $4\times 10^5$ steps, particles for intermediate values of $z$ orient preferentially along the $x$ axis. As the walls are approached, particles start to align to the $z$ direction. In this case, there is no disclination ring, but two disclination lines which emerge from the interfacial region. Fig. \ref{fig5} shows that the instantaneous ordering decays slowly, so we cannot preclude the possibility that the bridge relaxes to a similar configuration to the initial one. In order to check this possibility, we run longer simulations in absence of magnetic field and we found that the system seems to equilibrate to the new texture. In any case, this finding may indicate the presence of a locally stable bridge configuration in which particles orient preferentially along an axis perpendicular to the $z$ direction. Thus, at a second stage we applied a $z-$oriented magnetic field cycle. As $\lambda$ increases, particles orient mainly along the $z$ axis but, unlike the $D=30$ and $D=40$ cases for $a=1$, the bridge does not detach from the walls. When $\lambda$ reduces back to zero, Figs. \ref{fig6}c), \ref{fig6}f) and \ref{fig6}i) show that the bridge configuration returns to a similar configuration to the one after the initial $\lambda=0$ simulation, although no shearing is observed. However, examination of the nematic order parameter profiles and nematic director fields on different cross sections of the bridge shows that a disclination ring is formed in the equatorial ring, similar to the $D=20$ cases reported in Ref. \cite{Romero2023}, instead of the escaped radial configurations observed previously. Moreover, there are no indication of vertical disclination lines emerging from the interfacial region. In any case, we cannot preclude that longer simulations may transform this configuration into the starting one before applying any magnetic field. However, the similarities between the initial and final configurations clearly indicate that our procedure can be used to switch between locally stable configurations of the bridge. This method is reversible as the initial state can be recovered by applying an appropriate magnetic field cycle. 
\subsubsection{$D=30$}
\begin{figure}
\centering
\includegraphics[width=\textwidth]{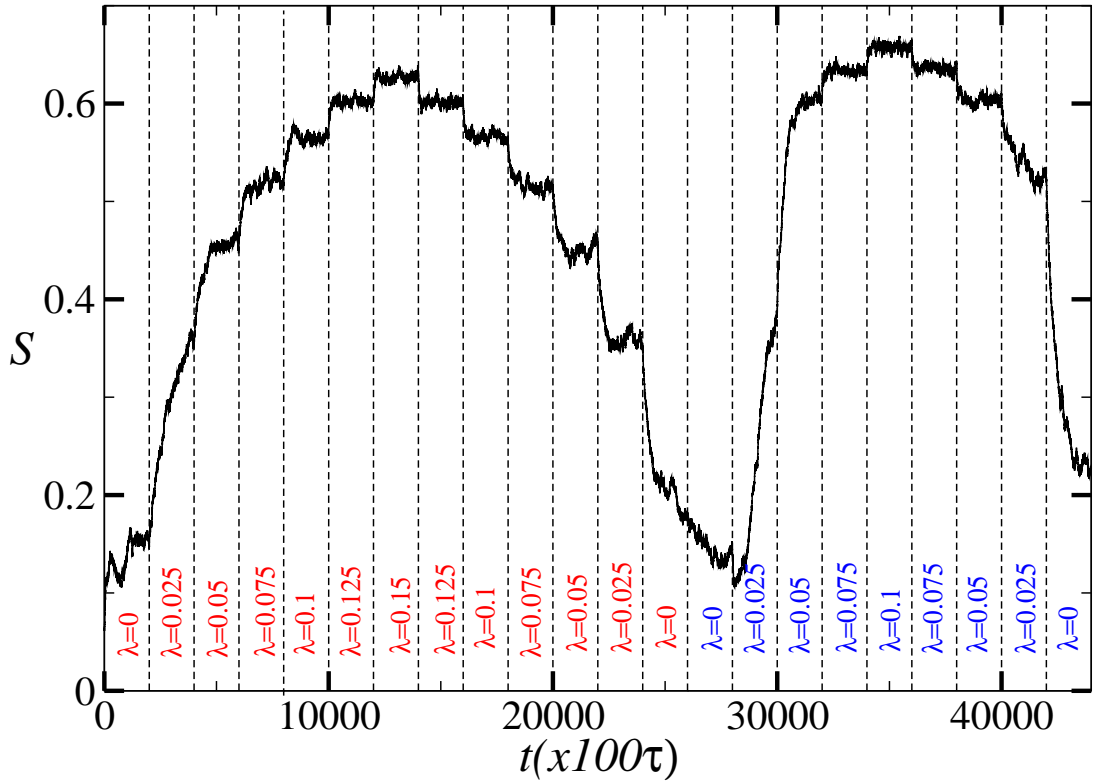} 
\caption{Time evolution of the instantaneous global nematic order parameter $S$ along the magnetic field application cycle for $D=30$ and $a=1.25$. The colour code for the values of $\lambda$ has the same meaning as in Fig. \ref{fig5}. 
} 
\label{fig8}
\end{figure}
\begin{figure}
\centering
\includegraphics{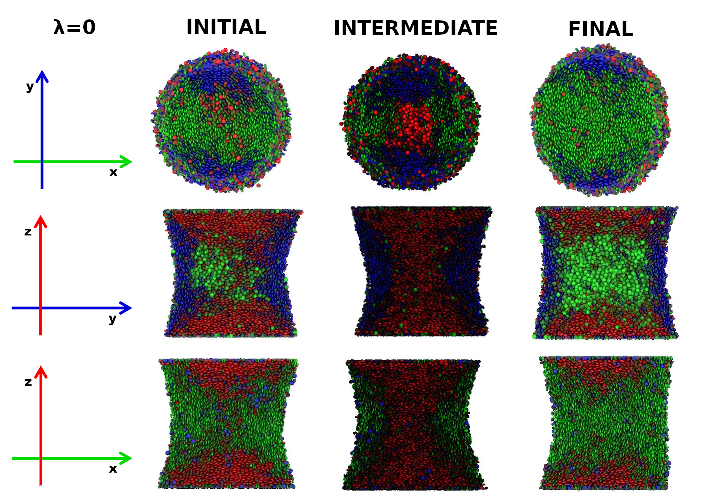} 
\caption{Snapshots of the $xy$, $yz$ and $xz$ cross sections of a bridge in a slit pore of two horizontal walls separated by a distance $D=30$ and $a=1.25$. The left column corresponds to the final configuration of the initial $\lambda=0$ simulation, middle column to the final configuration of the intermediate $\lambda=0$ simulation after the application of magnetic fields along the $z$ axis, and the right column to the final configuration of the last $\lambda=0$ simulation after the applicacion of magnetic fields along the $x$ axis. The colour code associated to particle orientations is the same as in Fig. \ref{fig1}.\label{fig9}}
\end{figure}
\begin{figure}
\centering
\subfloat{%
\resizebox*{5cm}{!}{\includegraphics{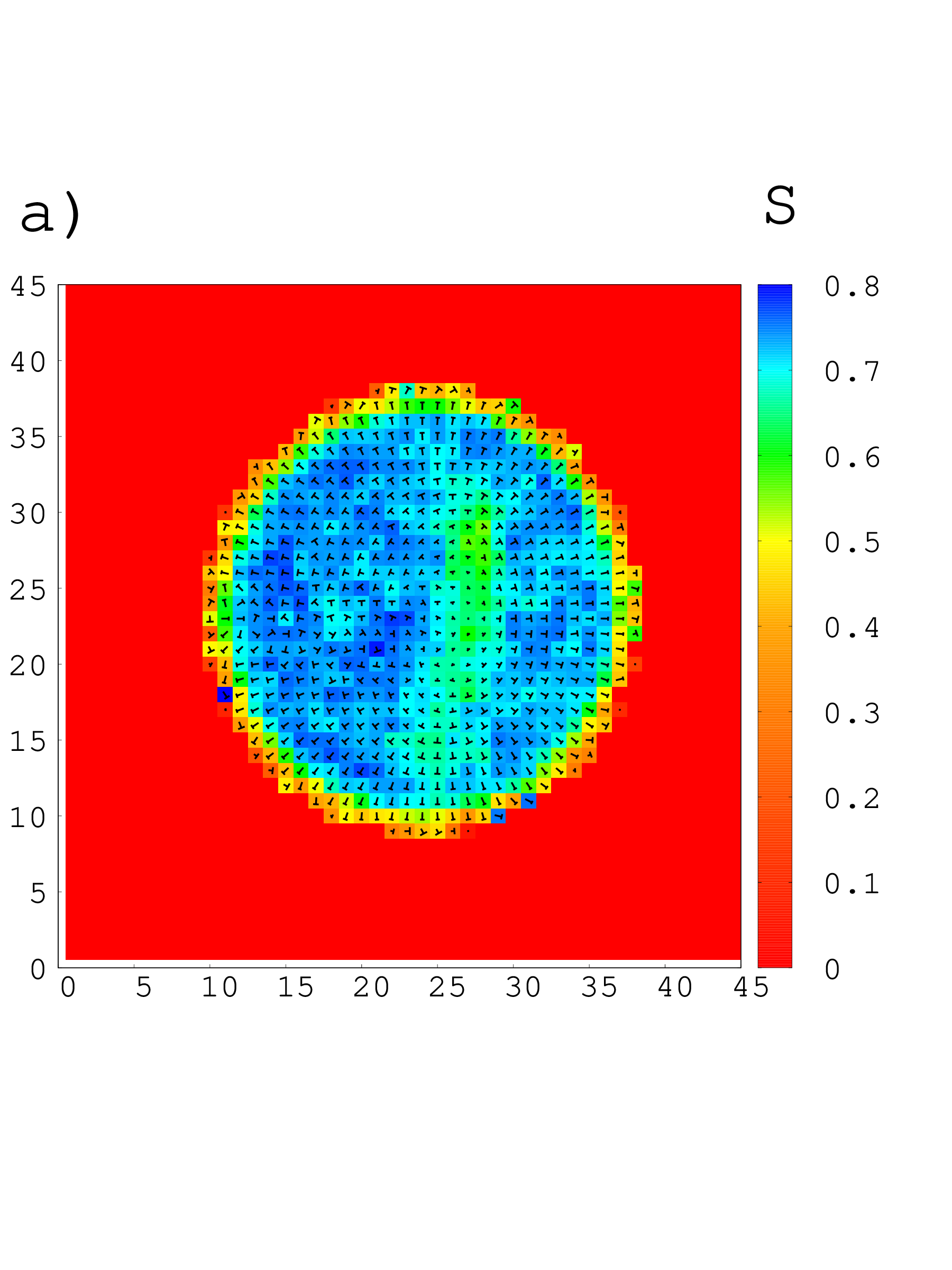}}}
\subfloat{%
\resizebox*{5cm}{!}{\includegraphics{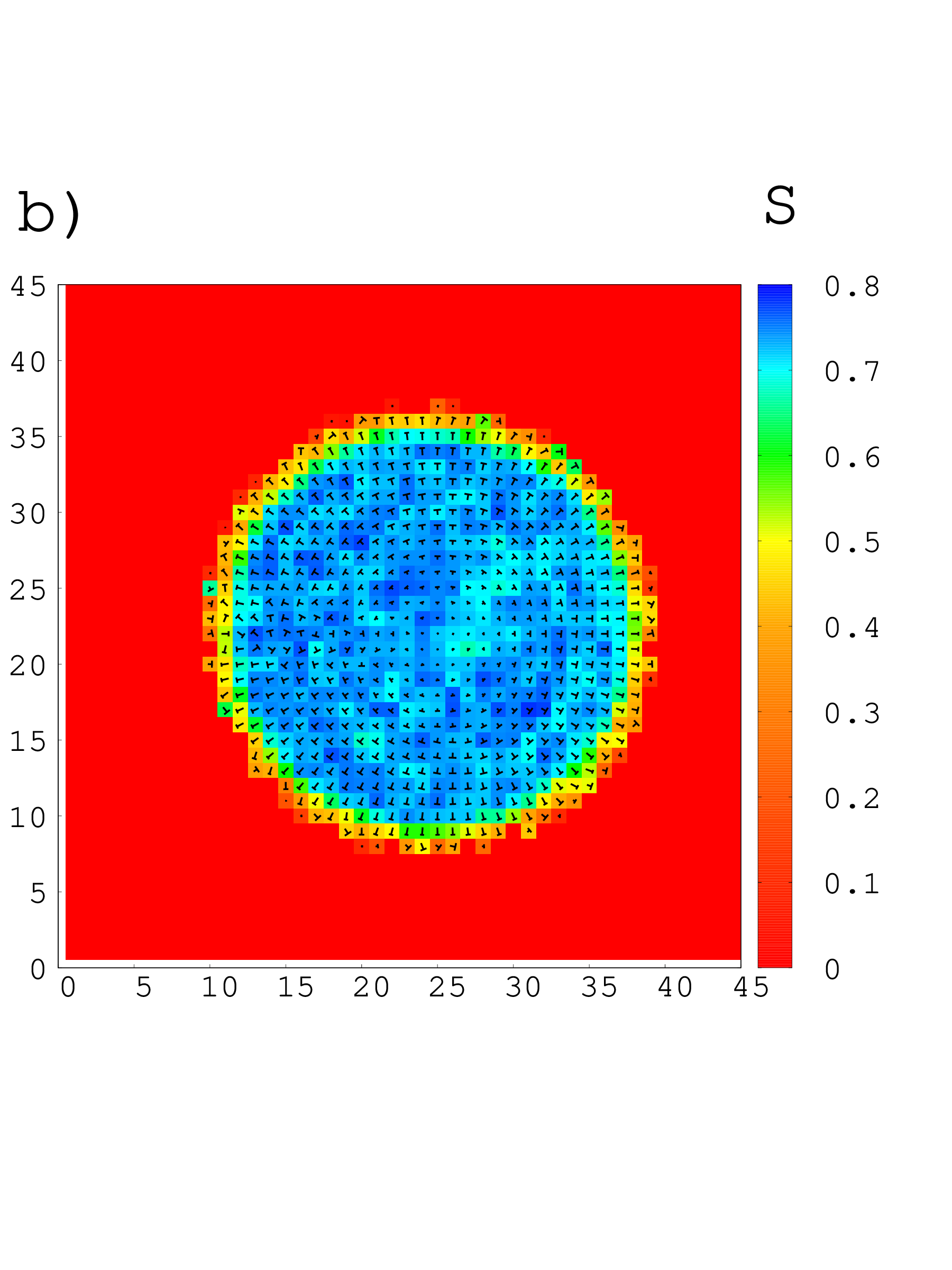}}}
\subfloat{%
\resizebox*{5cm}{!}{\includegraphics{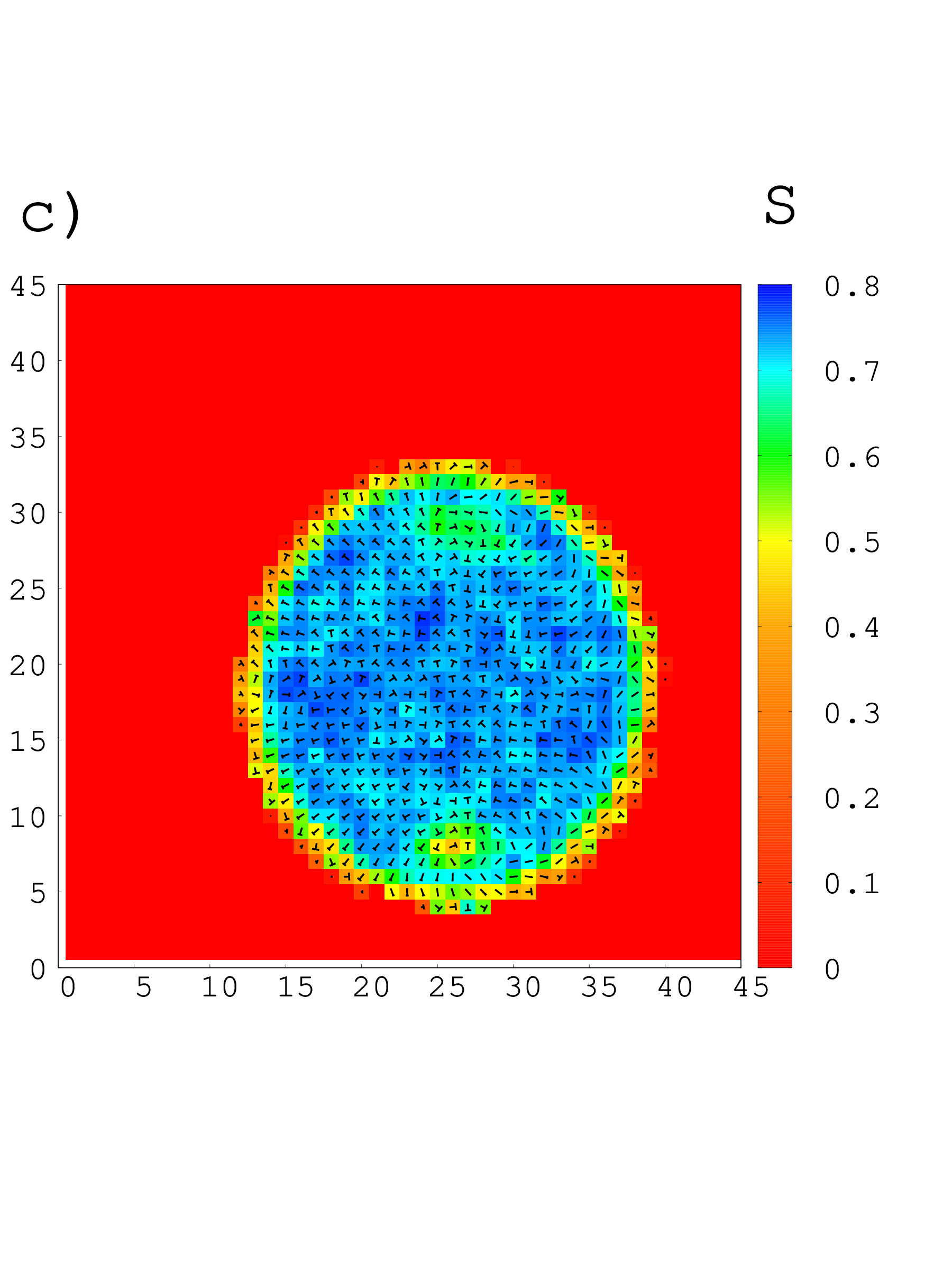}}}\vspace{-2.5cm}
\\
\subfloat{%
\resizebox*{5cm}{!}{\includegraphics{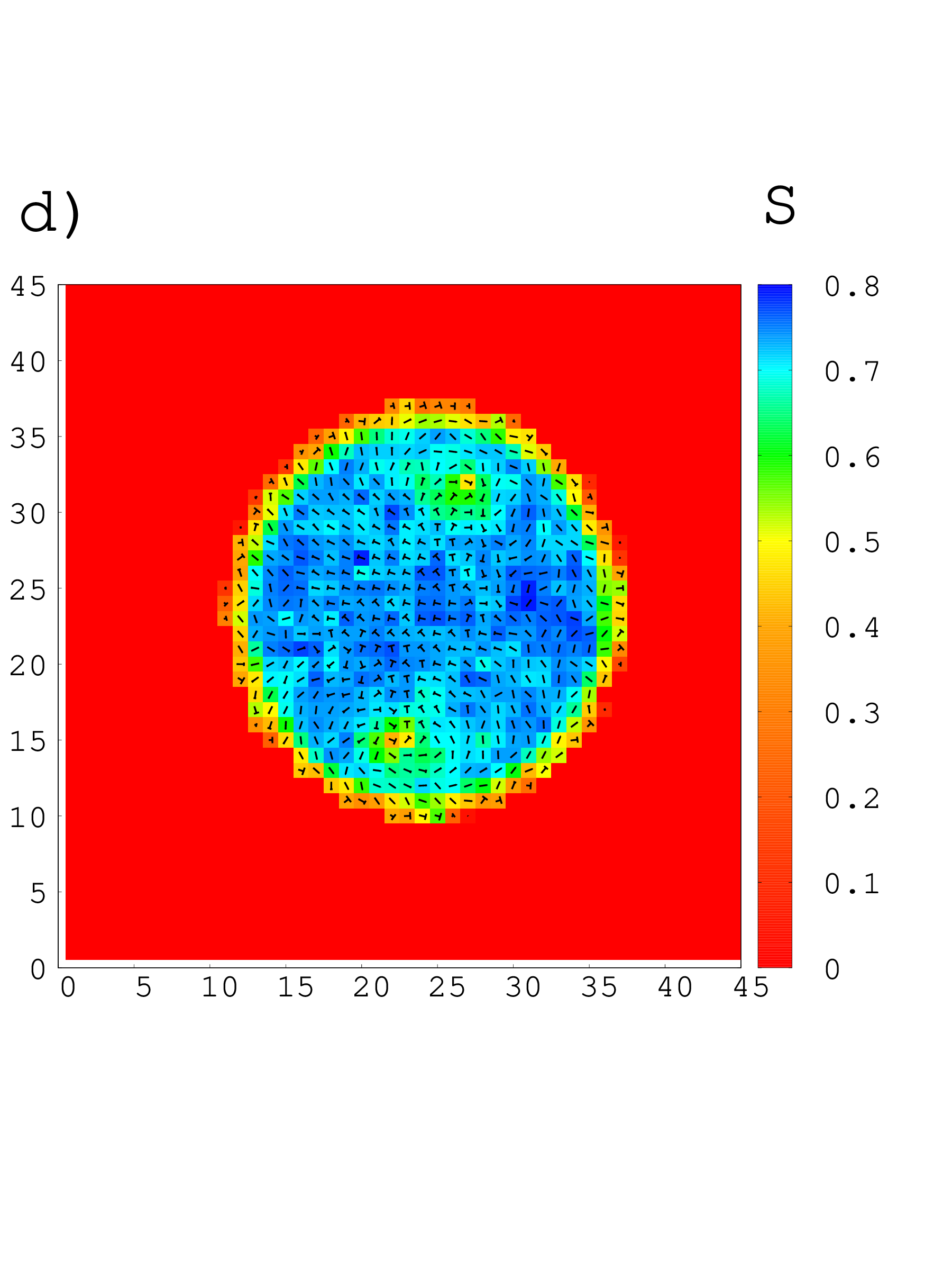}}}
\subfloat{%
\resizebox*{5cm}{!}{\includegraphics{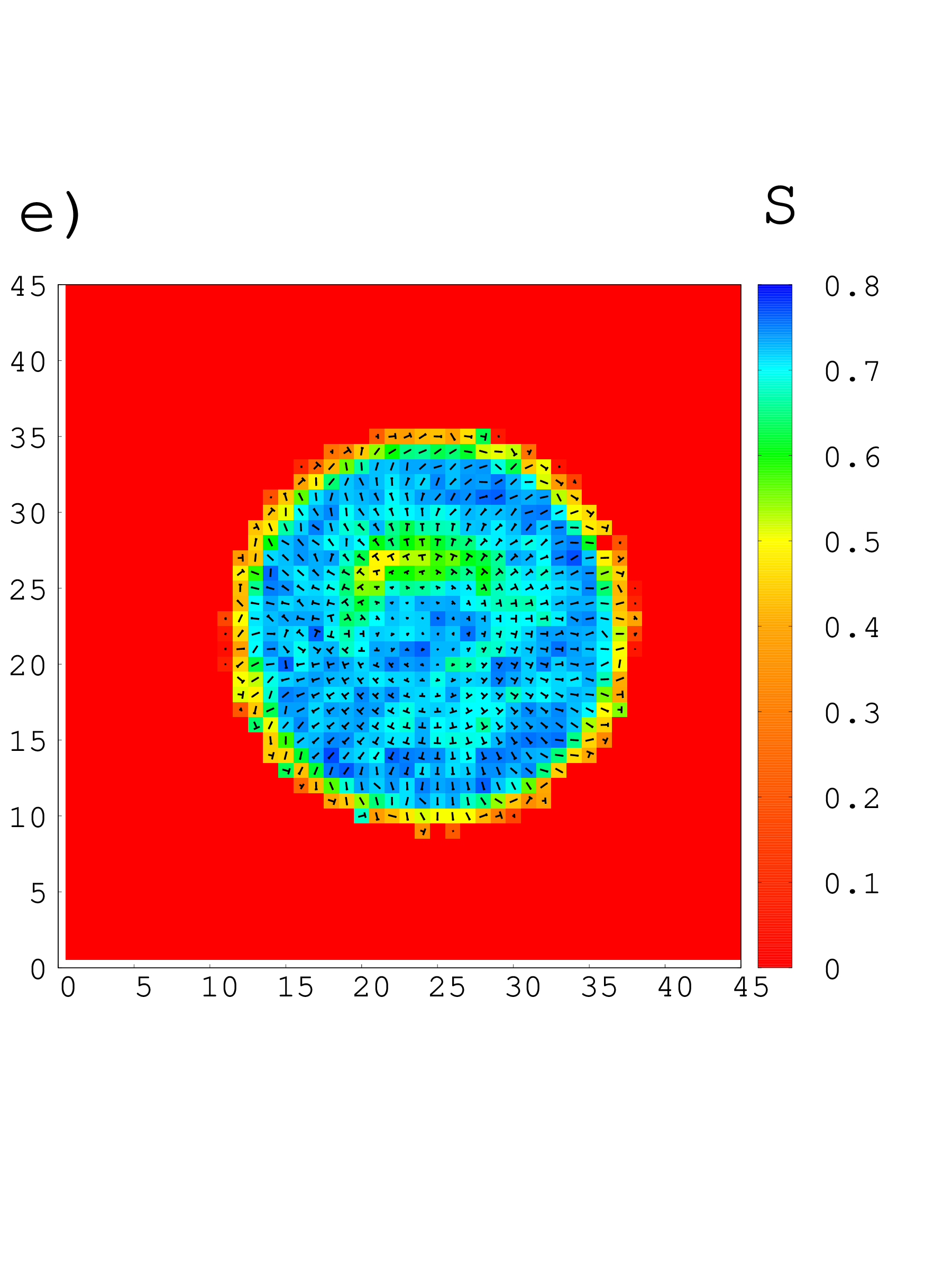}}}
\subfloat{%
\resizebox*{5cm}{!}{\includegraphics{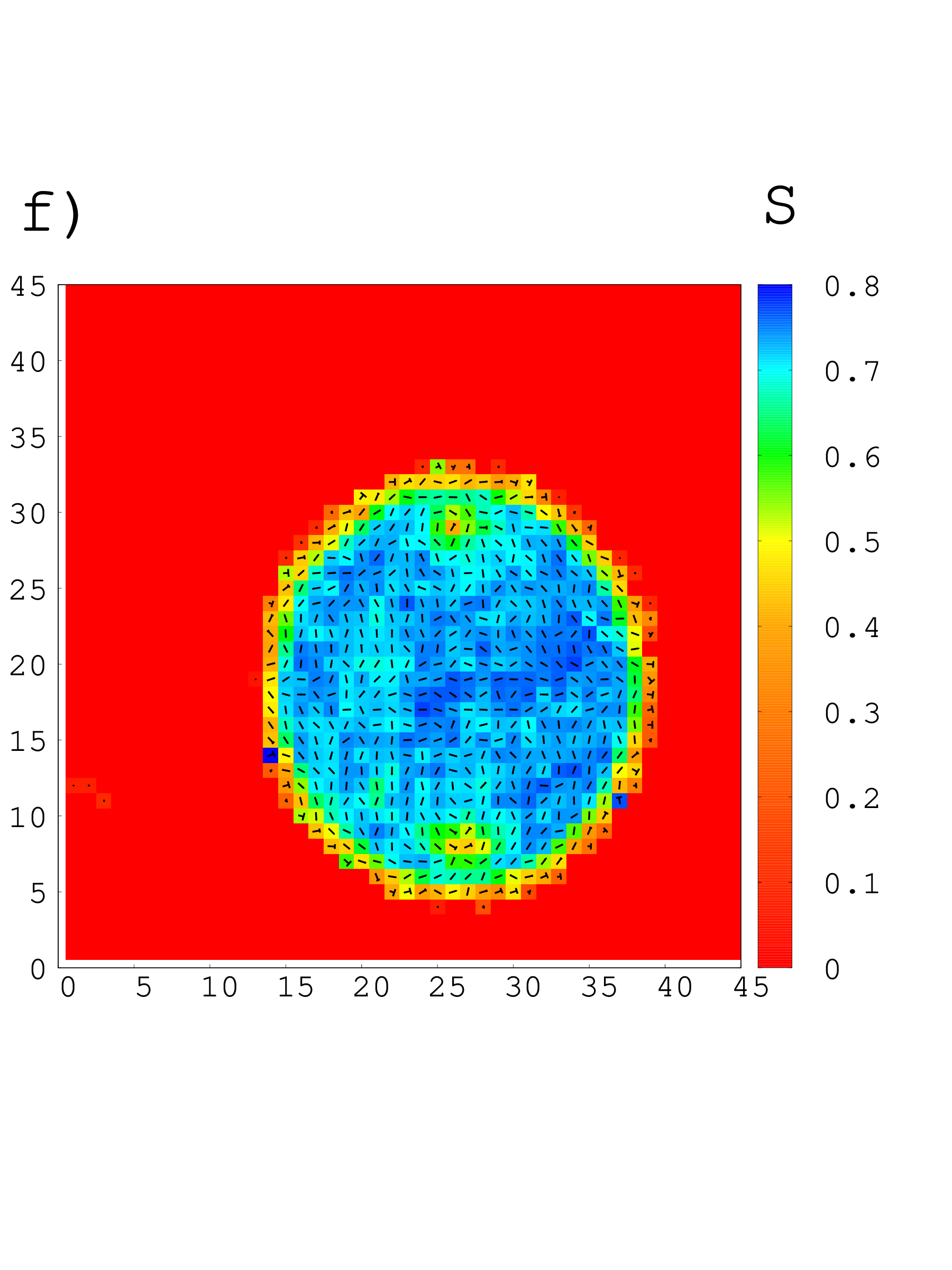}}}\vspace{-2.5cm}
\\
\subfloat{%
\resizebox*{5cm}{!}{\includegraphics{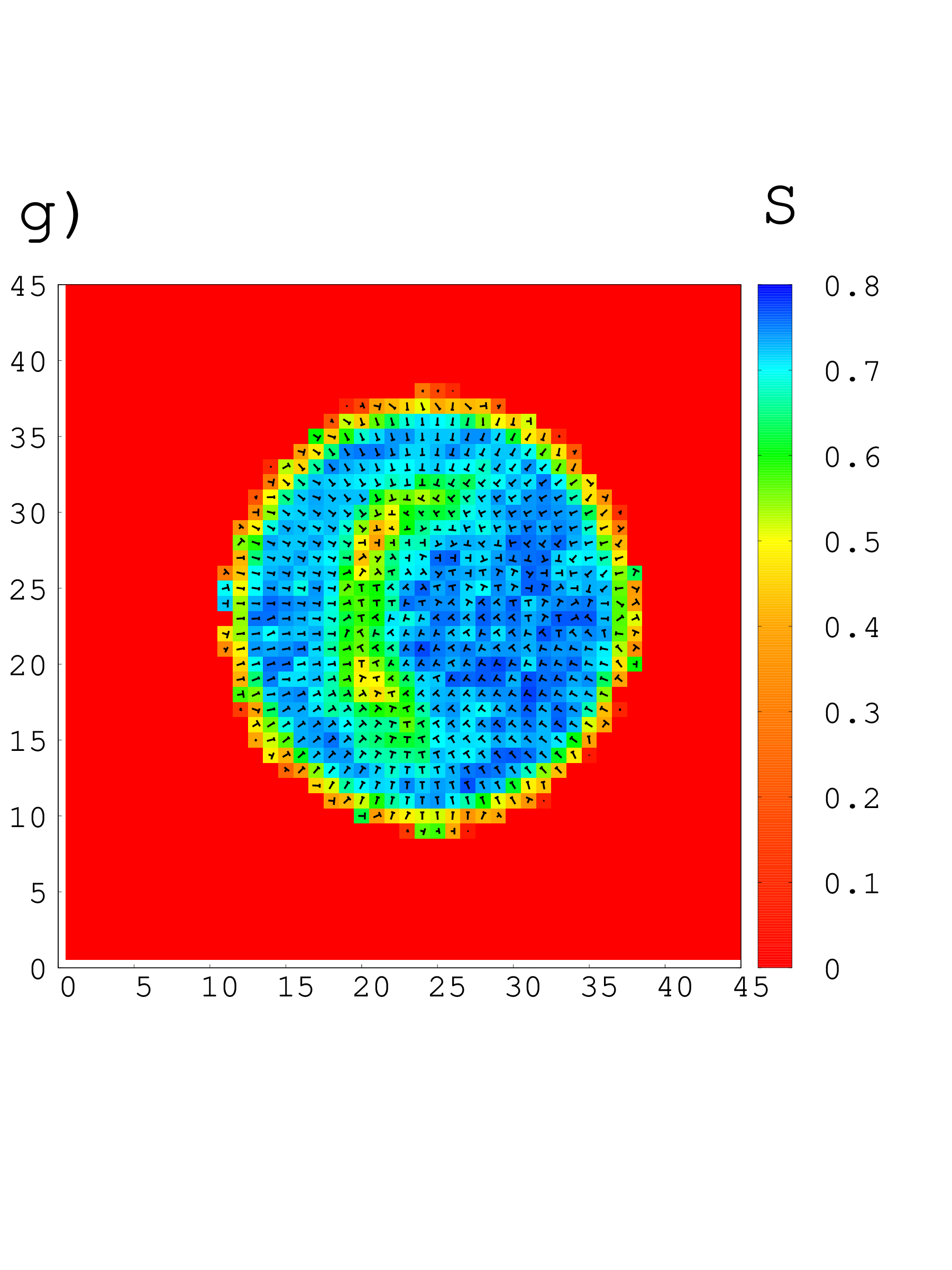}}}
\subfloat{%
\resizebox*{5cm}{!}{\includegraphics{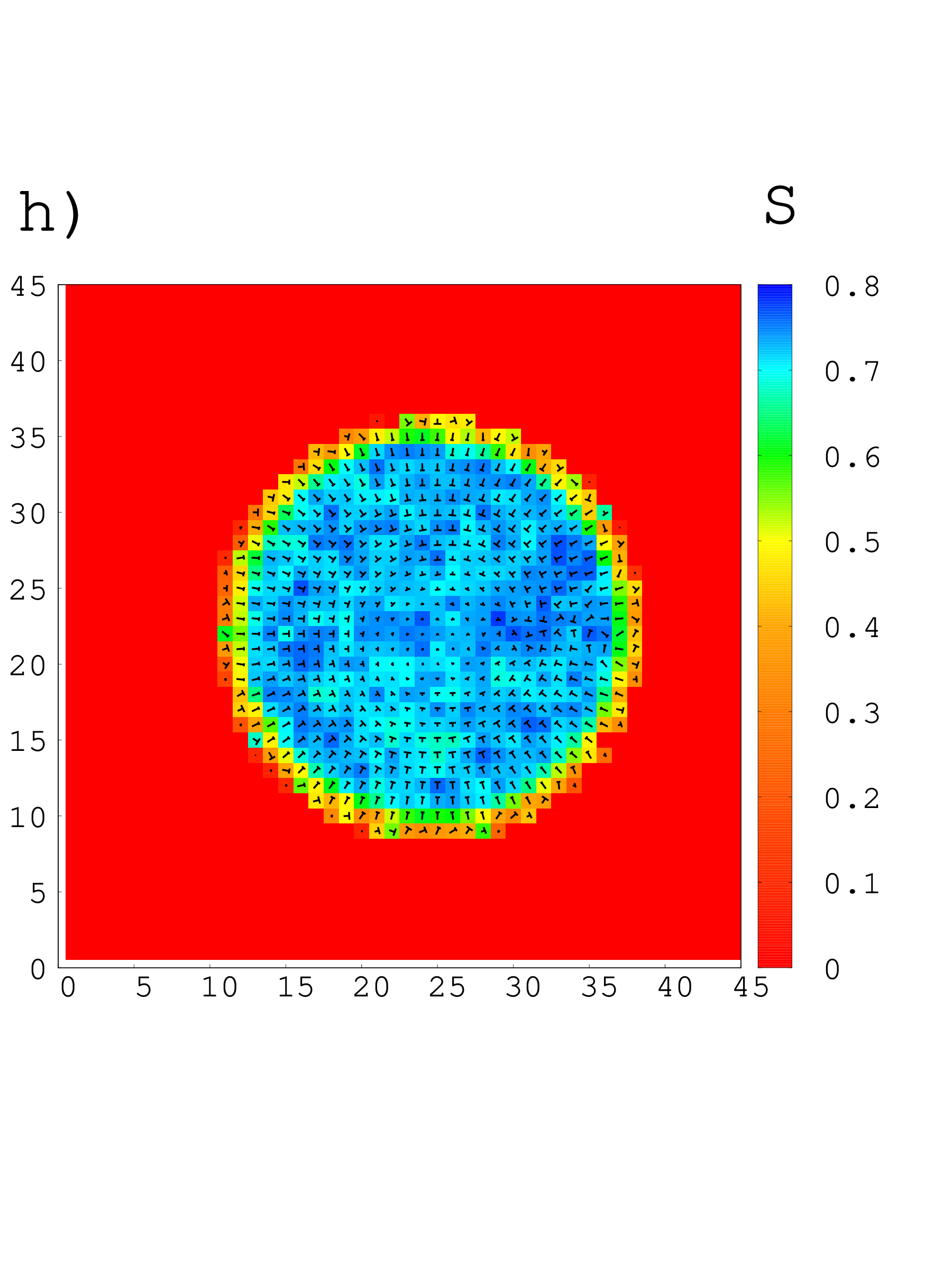}}}
\subfloat{%
\resizebox*{5cm}{!}{\includegraphics{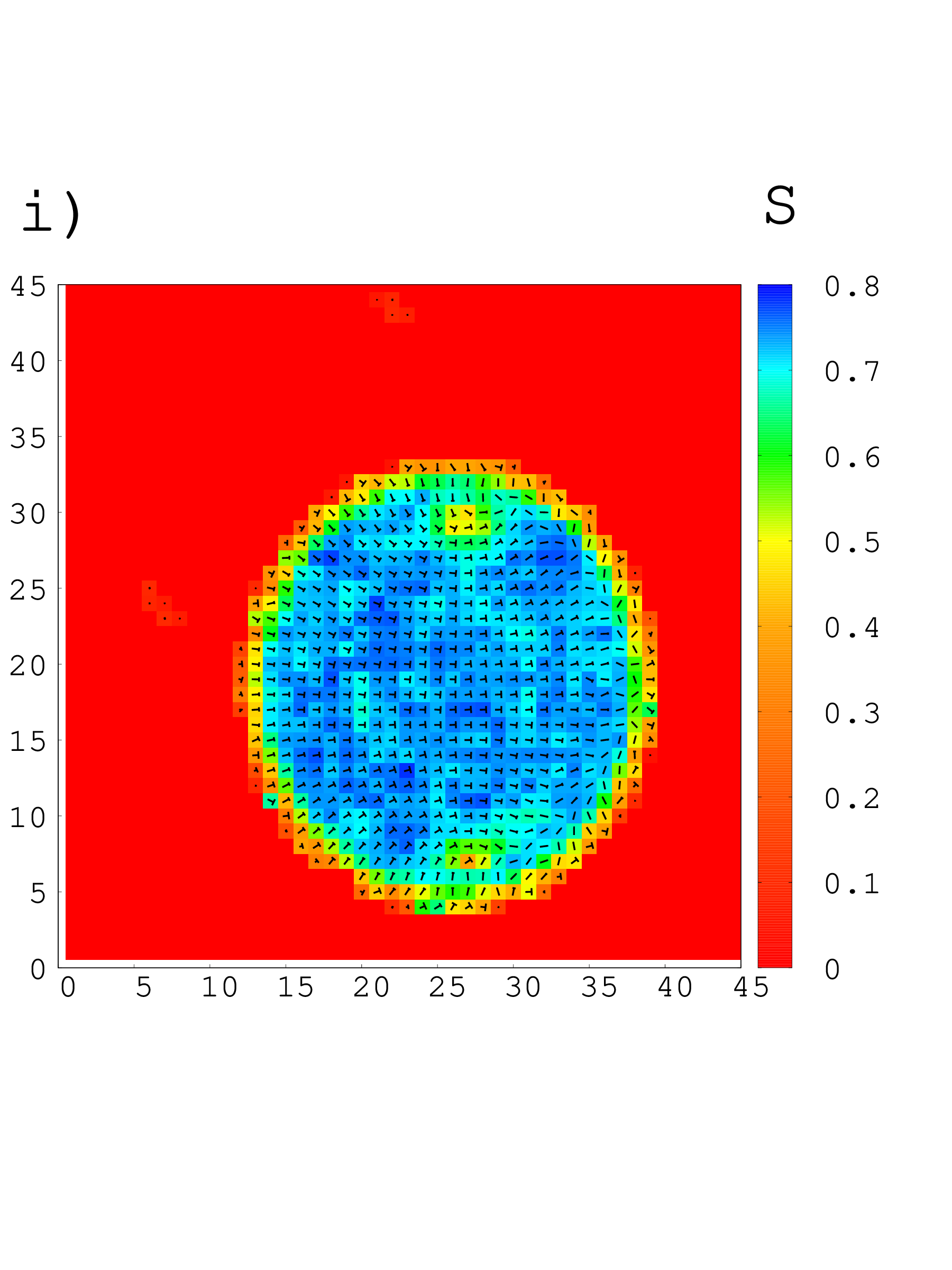}}}
\caption{Plots of the nematic order parameter $S$ profile (colour map) and nematic director field (nail representation) of a cross section of the bridge for $a=1.25$ and $D=30$ at heights $z=8$ (upper row), $z=15$ (middle row) and $z=22$ (lower row). Left column corresponds to the initial $\lambda=0$ simulation, central column to the intermediate $\lambda=0$ simulation and right column to the final $\lambda=0$ simulation.} \label{fig10}
\end{figure}
\begin{figure}
\centering
\subfloat{%
\resizebox*{5cm}{!}{\includegraphics{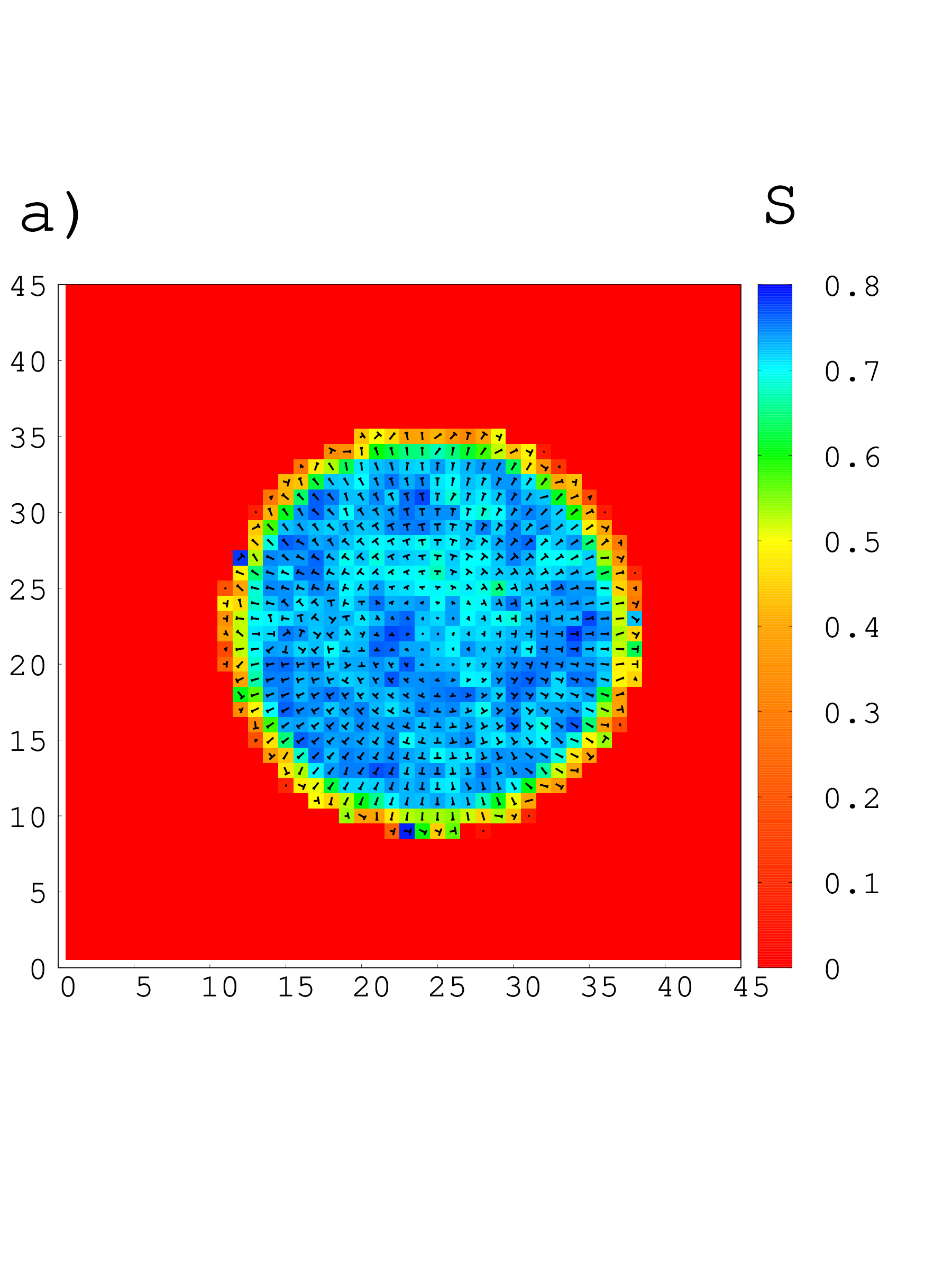}}}
\subfloat{%
\resizebox*{5cm}{!}{\includegraphics{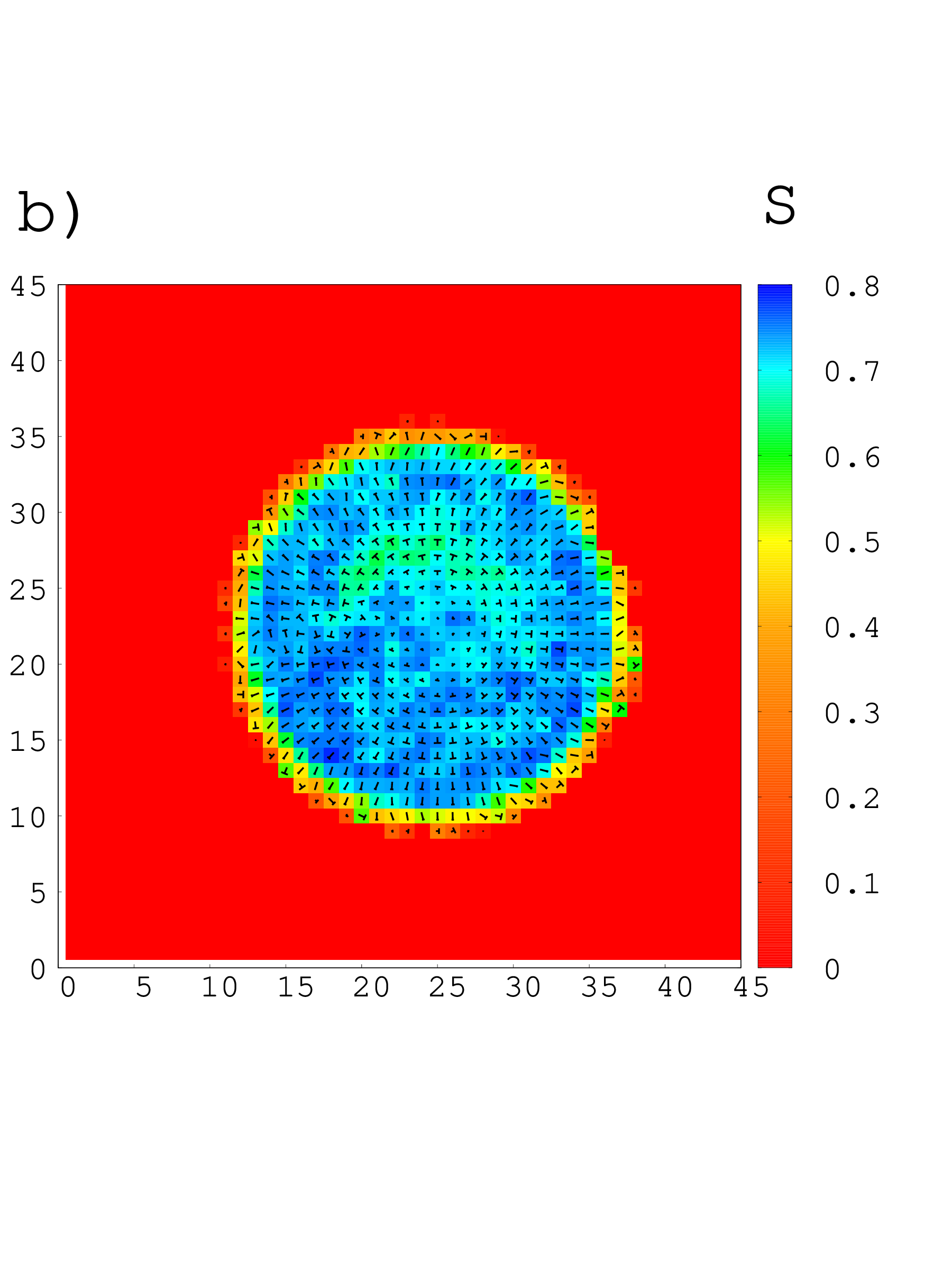}}}
\subfloat{%
\resizebox*{5cm}{!}{\includegraphics{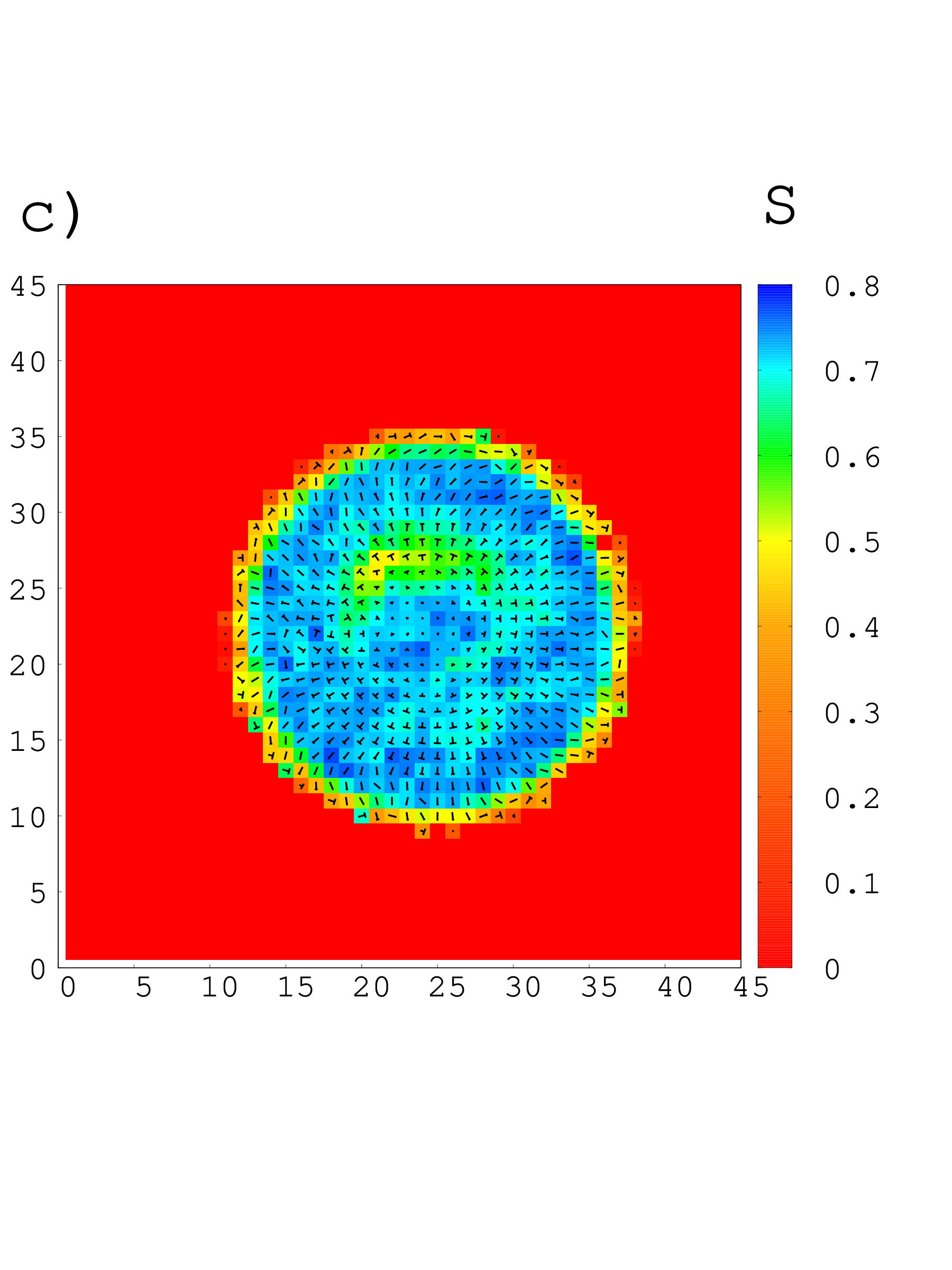}}}\vspace{-2.5cm}
\\
\subfloat{%
\resizebox*{5cm}{!}{\includegraphics{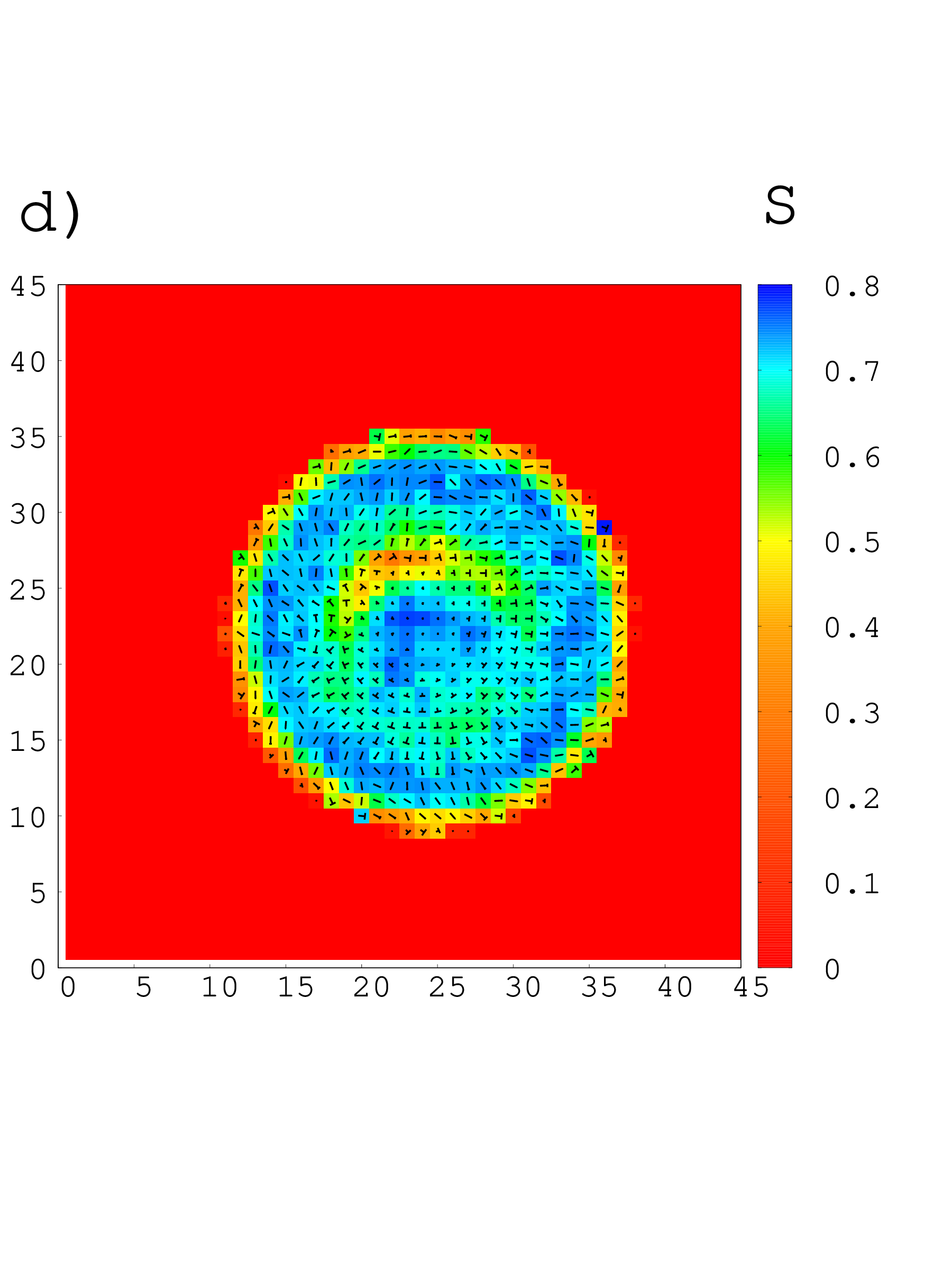}}}
\subfloat{%
\resizebox*{5cm}{!}{\includegraphics{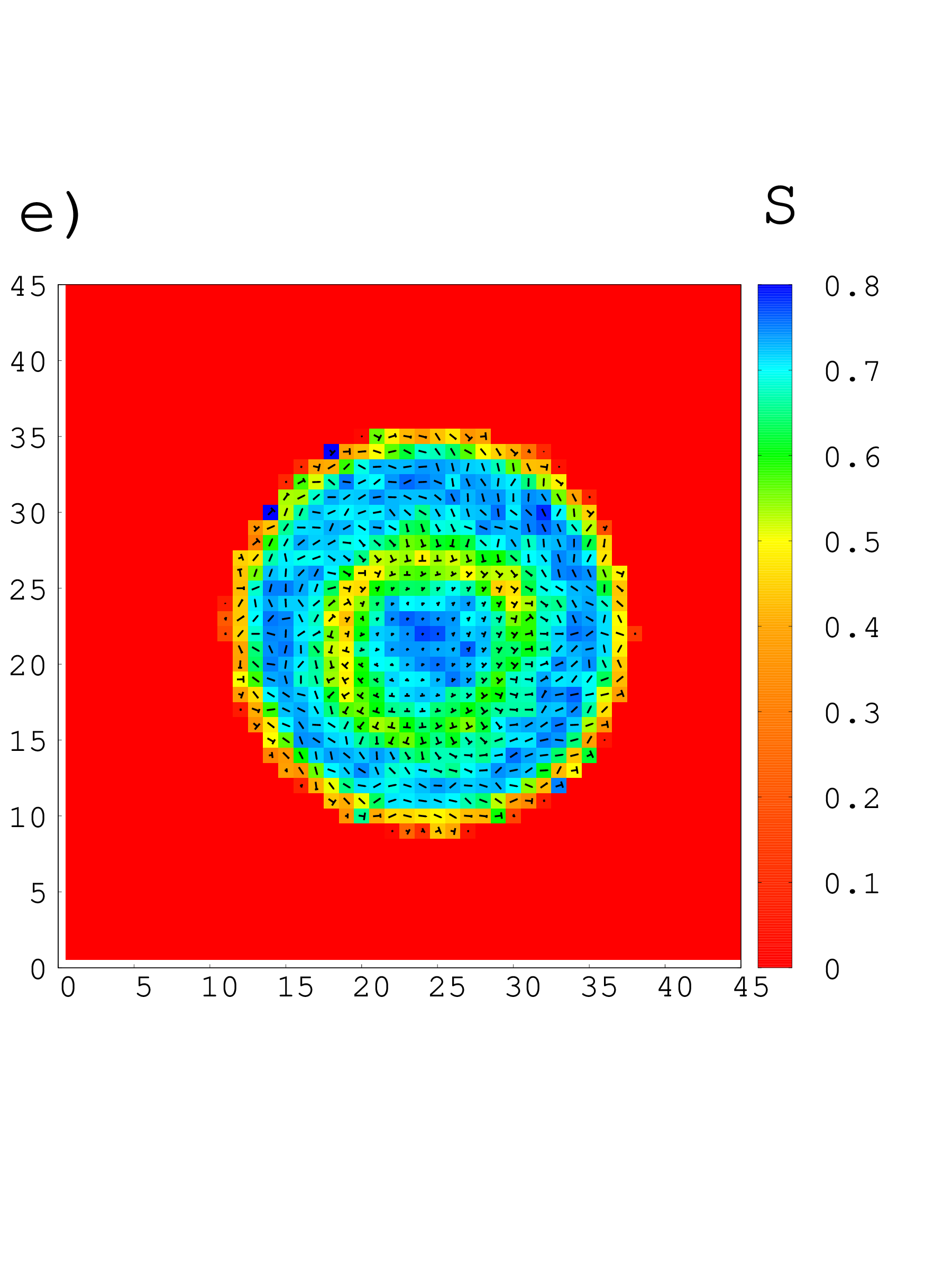}}}
\subfloat{%
\resizebox*{5cm}{!}{\includegraphics{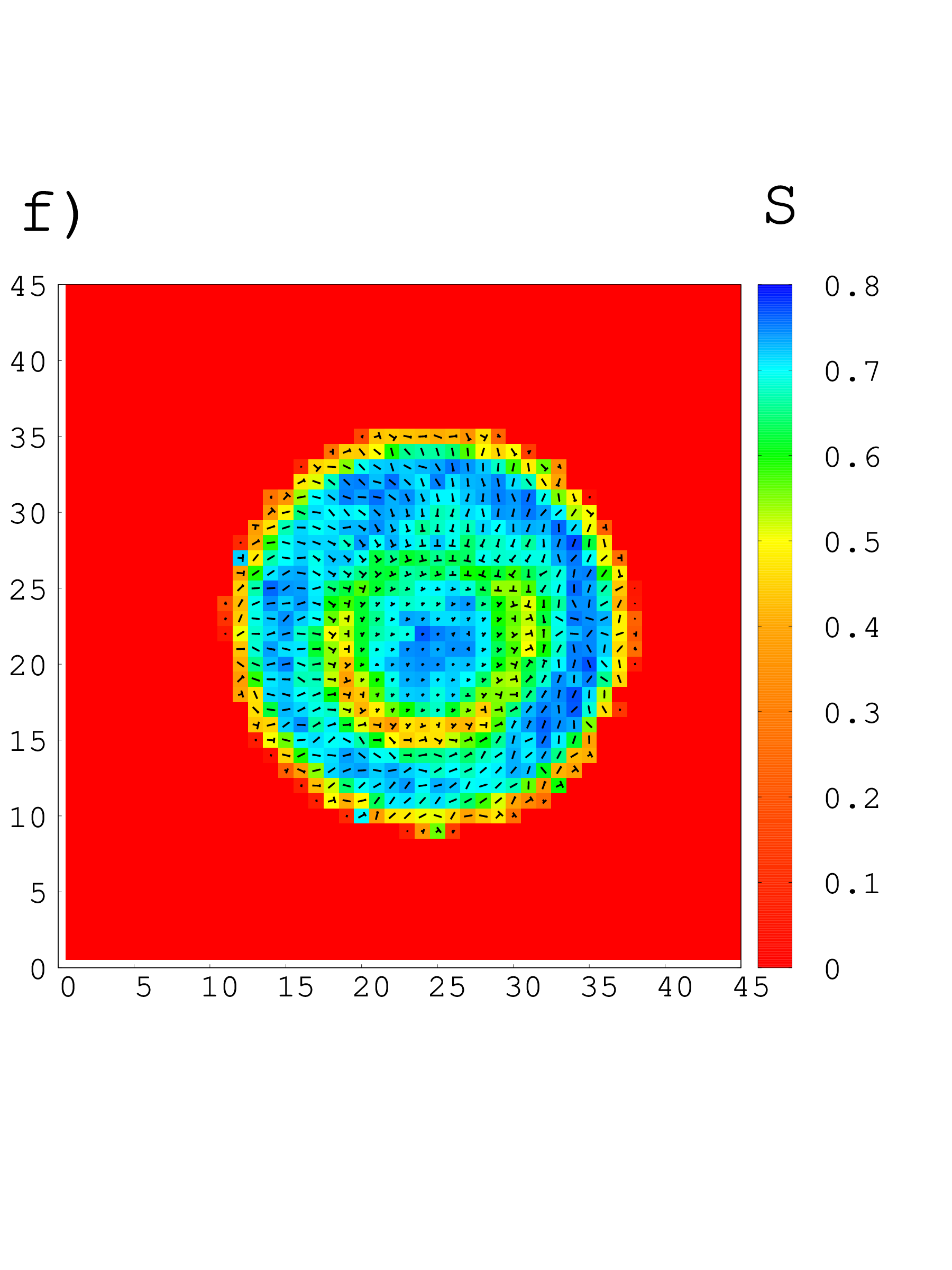}}}\vspace{-2.5cm}
\\
\subfloat{%
\resizebox*{5cm}{!}{\includegraphics{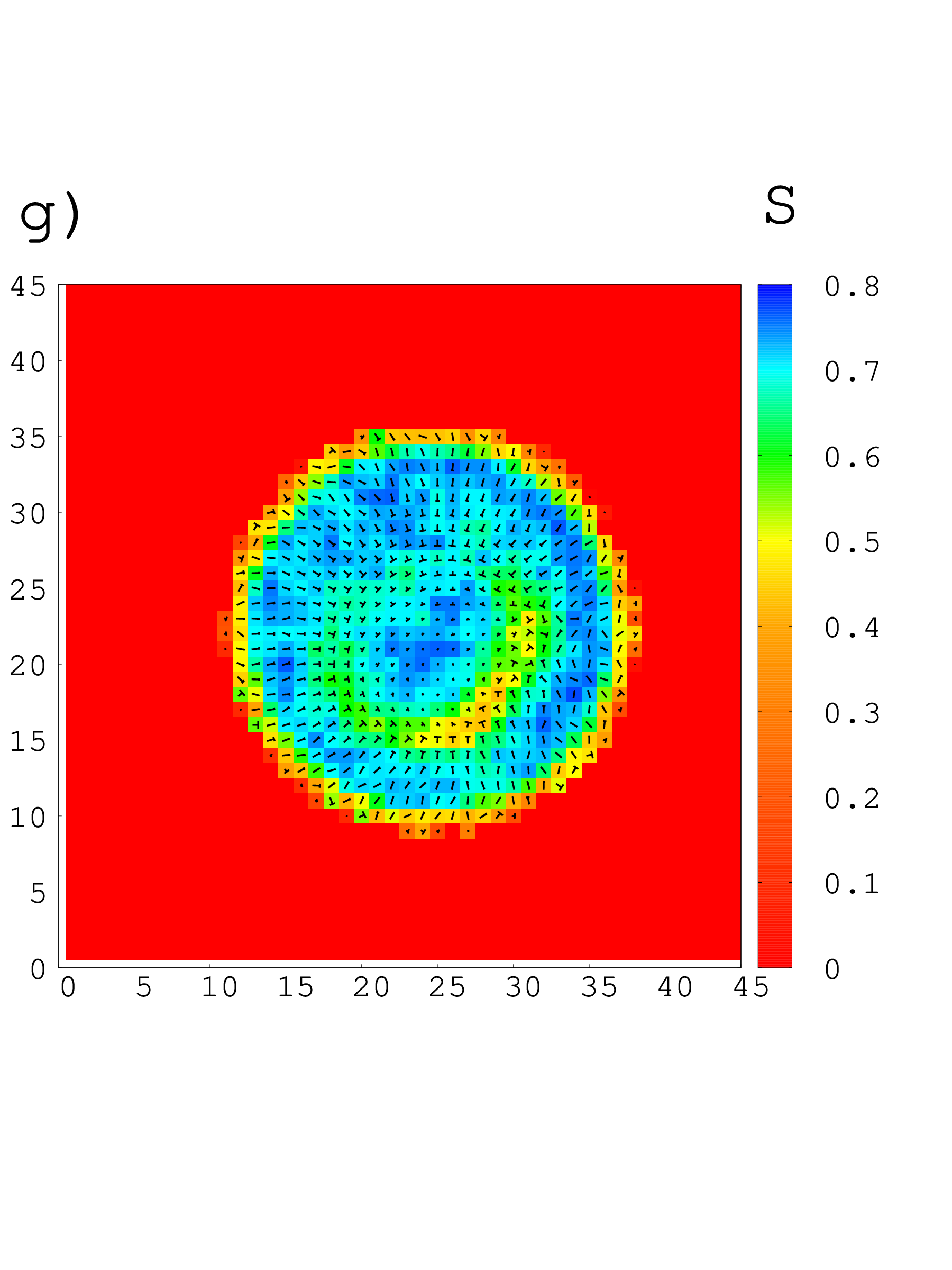}}}
\subfloat{%
\resizebox*{5cm}{!}{\includegraphics{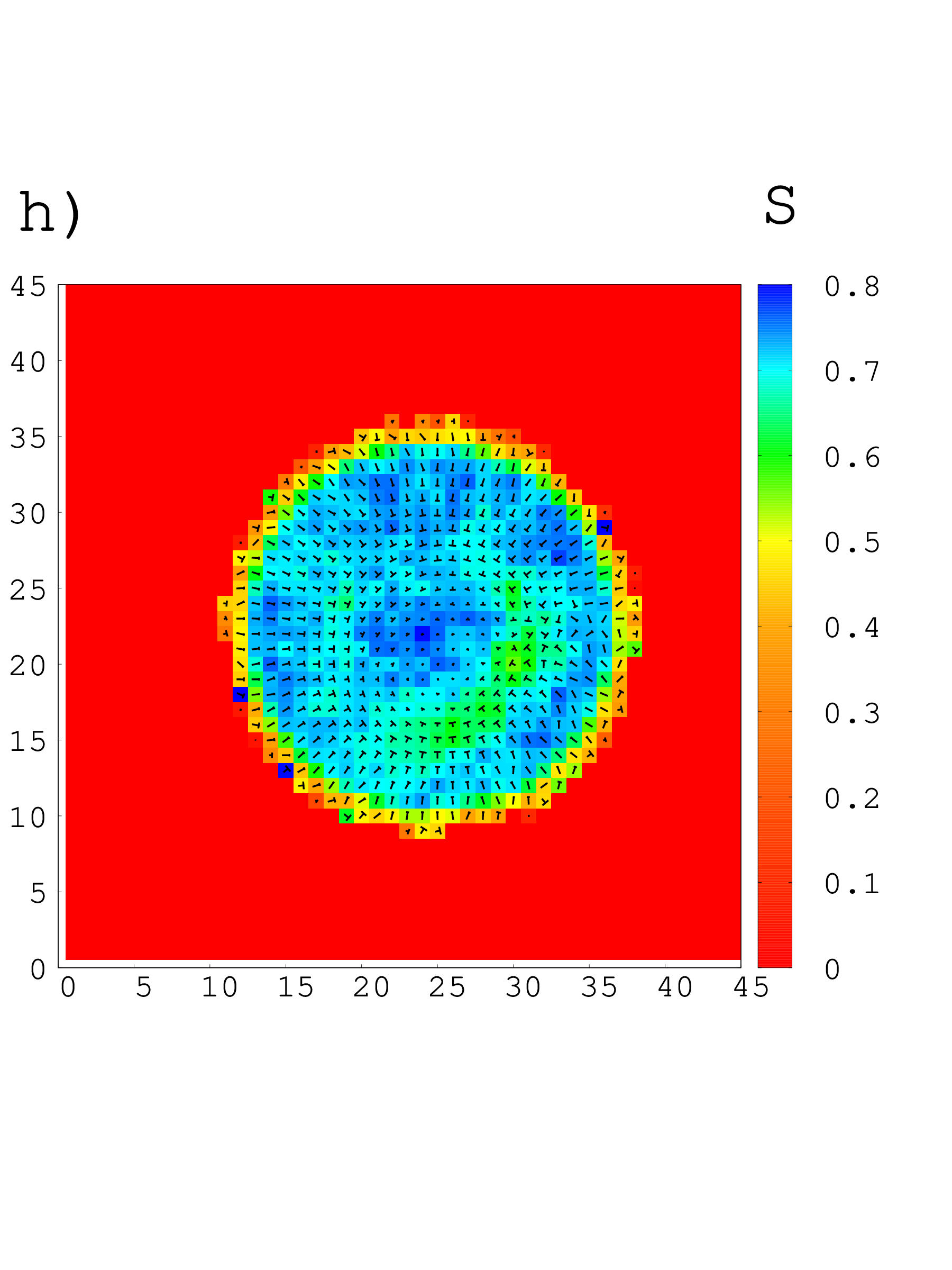}}}
\subfloat{%
\resizebox*{5cm}{!}{\includegraphics{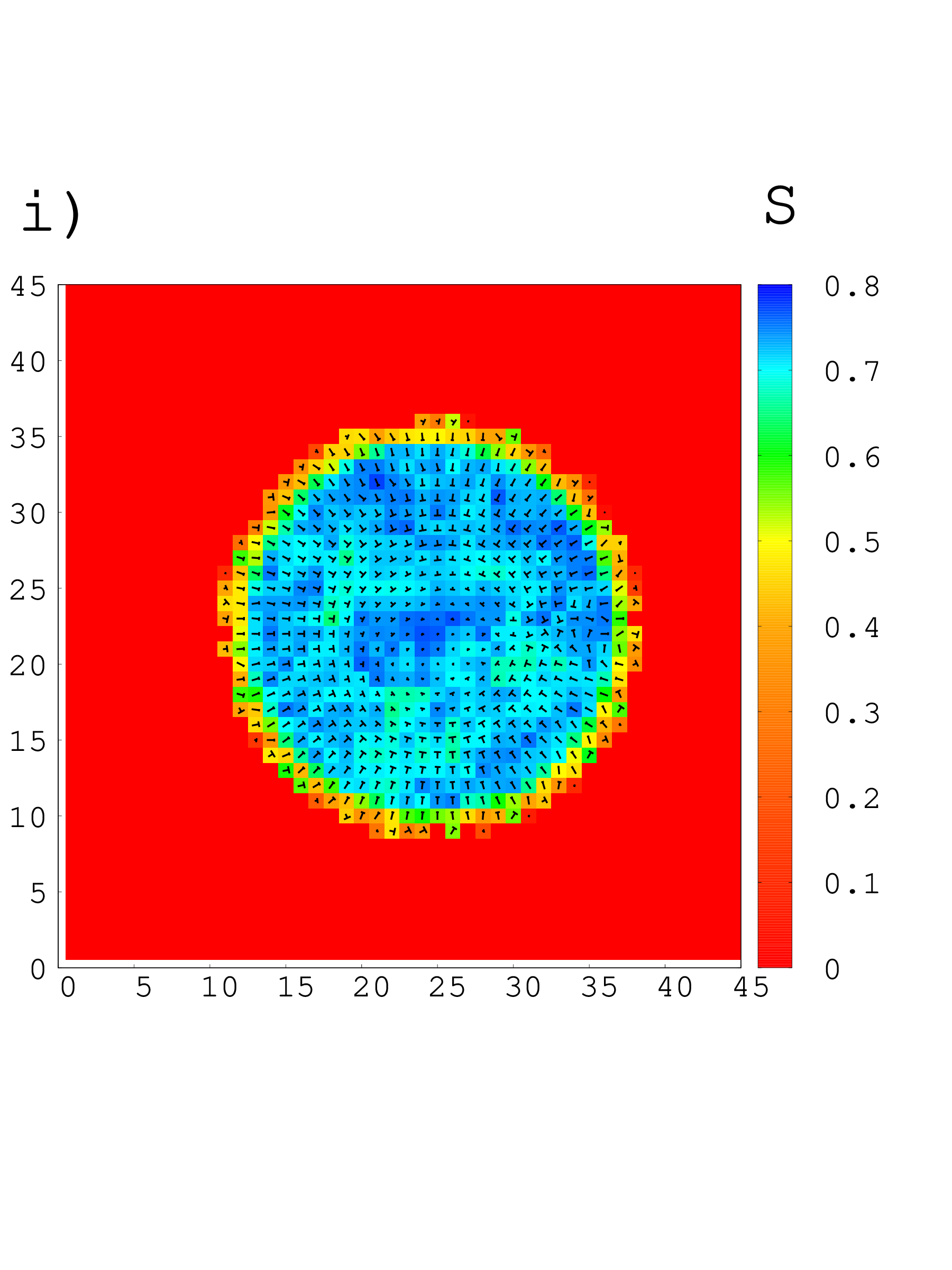}}}
\caption{Plots of the nematic order parameter $S$ profile (colour map) and nematic director field (nail representation) of different cross section of the bridge for $D=30$ and $a=1.25$ in the intermediate $\lambda=0$ simulations around the ring disclination line: a) $z=13$, b) $z=14$, c) $z=15$, d) $z=16$, e) $z=17$, f) $z=18$, g) $z=19$, h) $z=20$, and i) $z=21$.} \label{fig11}
\end{figure}

For $D=30$, in a first stage we apply magnetic fields along the $z-$direction. Fig. \ref{fig8} shows the evolution of the instantaneous nematic order parameter $S$ in a magnetic field application cycle. The initial $\lambda=0$ simulation leads to a similar configuration to the obtained for the $a=1$ case (compare left columns of Figs. \ref{fig3} and \ref{fig9}).  In this case, the two vertical disclination lines do not emerge from the interfacial region but they start approaching and merge close to the walls, as it is indicated by Figs. \ref{fig10}a) and \ref{fig10}c). We check that the merging occurs at heights $z\approx 8$ and $z\approx 24$. So, a vertical closed disclination ring is observed, as it was reported in Ref. \cite{Romero2023}. As the magnetic field is turned on for $\lambda=0.025$, particles orient quickly along the $z$ axis. As $\lambda$ increases, the degree of orientation along the $z$ axis is larger but the shape of the bridge is only slightly thinned on the equatorial plane. Thus, the bridge remains stable up to $\lambda=0.15$, unlike the $a=1$ case. The reduction of $\lambda$ reduces steadily the order parameter $S$ until that the magnetic field is turned off, in which a steep decrease indicate a major reorientation of particles in the bridge. For $4\times 10^5$ steps, $S$ changes slowly in a similar way as it did for the $D=20$ case (compare to Fig. \ref{fig5}). However, the configuration of the bridge is completely different from the initial one (see middle column of Fig. \ref{fig9}): now most particles orient along the $z$ axis except in the interfacial region. The order parameter profiles and nematic director fields shown in the middle column of Fig.~\ref{fig10} indicate that the nematic order parameter is uniform throughout the bridge core up to the interfacial region except near the equatorial plane ($z=15$). Around this height there is a extensive region in which $S$ is reduced with respect to the bulk value and that separates an inner region in which the nematic director is oriented along the $z$ axis from an outer shell away from the interfacial region where the nematic director is radial. This is an indication of the presence of a disclination ring on approximately the equatorial plane. This is confirmed by the analysis of the orientational order profiles and director fields at different cross-sections of the bridge from $z=13$ to $z=21$ (see Fig. \ref{fig11}). Away from this region, the nematic director field shows a escaped radial configuration, as it can be seen in Figs. \ref{fig10}b) and \ref{fig10}c). Thus, the application of the magnetic field cycle can again switch the configuration of the bridge. We confirmed at least the local stability of this configuration by additional longer zero-field simulations, in which the bridge apparently equilibrated without major changes in its texture.

In order to see if this change can be reverted, a second magnetic field cycle is applied, in which now the magnetic field is oriented along the $x$ axis. As it happened in the $D=20$ case, Fig. \ref{fig8} shows, as the magnetic field is turned on, major particle reorientations occur  for $\lambda=0.025$ and $\lambda=0.05$. In addition, the bridge is stretched along the $y$ axis in a similar way as for the previous $D=20$ cases. However, the bridge remains stable for $\lambda=0.1$. As the magnetic field is turned off, the nematic ordering reduces steadily until the magnetic field vanishes, in which again there is a steep decrease. The final snapshots of the bridge configuration (right column of Fig. \ref{fig9}) show a similar orientational organization to the final snapshot of the initial $\lambda=0$ simulation. However, the nematic order parameter $S$ profile and the nematic director field shows some differences. For example, right column of Fig. \ref{fig10} indicate that there are two vertical disclination lines which, instead of merging, emerge and disappear in the interfacial region. Nevertheless, we can safely state that the bridge returned to the initial state.
\subsubsection{$D=40$}
\begin{figure}
\centering
\includegraphics{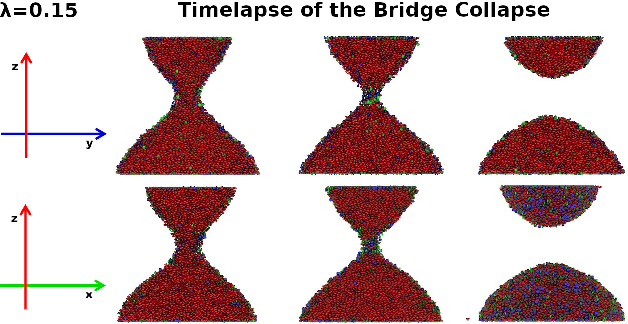} 
\caption{Time evolution of the bridge collapse for $D=40$, $a=1.25$ and $\lambda=0.15$. The $xz$ and $yz$ central sections of the bridge snapshots are shown for the time step $t=1.5\times 10^5$ (left column), $t=1.75\times 10^5$ (middle column) and $t=2\times 10^5$ (right column).
} 
\label{fig12}
\end{figure}

Finally we applied a $z-$oriented magnetic field cycle to the $D=40$ case. In the initial $\lambda=0$ case, the final configuration resembles the one obtained from Monte Carlo simulation in Ref. \cite{Romero2023}. Under the application of a magnetic field along the $z$ direction, particles reorient vertically, but in addition the bridge is squeezed on the equatorial plane. If the magnetic field is strong enough, the bridge breaks down from the middle part and two droplets, each in contact with a different wall, are formed. Fig. \ref{fig12} shows this breakdown for $\lambda=0.15$. As it happened to the $D=30$ and $D=40$ cases for $a=1$, this process cannot be reverted by switching off the magnetic field and thus it is irreversible.

\section{Conclusions}

In this paper we report preliminary molecular dynamic study of nematic nanobridges between slit pores of width $D$ under the effect of uniform magnetic fields which are modulated in time. We have considered the cases reported in Ref. \cite{Romero2023}. For adimensional wall-particles interaction parameter $a=1.25$ and $D=20$, under the application of magnetic field along the $x$ axis we successfully switched from a axisymmetric nanobridge conformation in which particles orient preferentially along the $z$ axis to a non-symmetrical bridge configuration in which particles orient along the $x$ axis. The bridge conformation was reverted to the original one by application of a $z-$oriented magnetic field cycle. However, this switching is not observed for $a=1$, which may indicate that there is no locally stable non-symmetrical bridge configuration under these conditions. However, this issue needs further analysis to be confirmed.

For $a=1.25$ and $D=30$, the initial non-symmetrical nanobridge conformation characterized by a disclination ring on a vertical plane is switched to a axisymmetric configuration with a disclination ring on the equatorial nanobridge plane under a $z-$oriented magnetic field cycle, which can be reverted to the initial state by an $x-$oriented magnetic field cycle. However, if the parameter $a$ is reduced to $1$, the bridge destabilizes for large magnetic field intensities as it detaches from one of the walls of the slit pore, forming a single droplet in contact with the other wall. This process cannot be reverted by reducing the intensity field. A similar behaviour is observed for $D=40$ for both $a=1$ and $1.25$. For the latter, the nanobridge instead splits into two droplets on each wall of the pore. These instabilities are related to the shape deformation of the bridge under large magnetic field intensities when field is applied along the $z$ axis. The dependence of the onset of this instability on $a$ and $D$ is an important aspect to study in future, as it may limit the applicability of the magnetic-field-induced switching between different nanobridge configurations. 

\section*{Acknowledgements}

This paper is dedicated to Prof. J. L. Fern\'andez Abascal and to the memory of the Prof. L. F. Rull, whose achievements contributed significantly to the promotion of the Molecular Simulation community in Spain. In particular, J.M.R.-E. is deeply indebted to Prof. Rull, whose guidance and support were illuminating to him, as for many generations of students and colleagues. His outstanding contributions in the field of computer simulations, specially in the field of liquid crystals, are a memorial to a life committed to science. 

\section*{Disclosure statement}

The authors report there are no competing interests to declare.

\section*{Funding}

This work was supported by the Ministerio de Ciencia, Innovaci\'on y Universidades (Spain) under Grant numbers  PID2021-126348NB-I00 and PID2022-140061OB-100. P. R.-L. also acknowledges the Ministerio de Ciencia, Innovaci\'on y Universidades (Spain) for a predoctoral FPI Grant No. PREP2022-000273.

\end{document}